\documentclass[letterpaper]{article} 
\usepackage{aaai2027} 
\usepackage[hyphens]{url} 
\usepackage{graphicx} 
\usepackage{natbib} 
\usepackage{caption} 
\usepackage{booktabs}
\usepackage{amsmath}
\usepackage{amssymb}

\usepackage{array}
\usepackage{adjustbox}

\usepackage{multirow}

\title{Structural Guidance for Unified Joint Demosaicing and Denoising}

\author{
Qixin Zheng\textsuperscript{1},
Ping Chen\textsuperscript{1},
Qiangqiang Shen\textsuperscript{1},
Haijin Zeng\textsuperscript{1}\corresponding
}

\affiliations{
\textsuperscript{1}\textit{Harbin Institute of Technology (Shenzhen)}, Shenzhen, China\\

haijin.zeng2018@gmail.com\\[2pt]

}

\newcommand{\FigureOrPlaceholder}[4]{%
\IfFileExists{#1}{%
\includegraphics[width=#2,height=#3,keepaspectratio]{#1}%
}{%
\fbox{\parbox[c][#3][c]{\dimexpr#2-2\fboxsep-2\fboxrule\relax}{\centering\small #4\par\vspace{1mm}\scriptsize Replace with: \texttt{\detokenize{#1}}}}%
}%
}

\nocopyright
\begin{document}

\maketitle

\begin{abstract}

{
Joint demosaicing and denoising is a fundamental step in camera image signal processing, yet remains challenging because different Bayer-like color filter arrays (CFAs) and sensor noise jointly corrupt both color sampling and image content. Existing unified restoration networks explicitly model CFA geometry but are still driven primarily by pixel-level supervision, making them prone to structural degradation around edges, repetitive textures, and moiré patterns where local evidence is unreliable. We attribute this limitation partly to the absence of explicit structural guidance beyond pixel-level reconstruction supervision. Motivated by this observation, we propose a structural-guided unified restoration framework that injects pretrained structural knowledge into CFA-aware image restoration. Our model receives a unified five-channel observation consisting of the raw mosaic, CFA masks, and a noise-level map. A SwinIR restoration branch reconstructs pixel details under CFA-conditioned modulation, while a parallel structural reasoning branch extracts complementary structural cues from a sparse pseudo-RGB observation. To bridge the substantial domain gap between sparse noisy sensor data and the natural-image pretraining domain of the structural encoder, we introduce a lightweight trainable adapter before residually fusing structural and restoration features. A shared decoder jointly predicts the restored RGB image and an auxiliary clean mosaic, providing supervision in both image and sensor domains. Extensive experiments across multiple CFA patterns and noise levels demonstrate consistent improvements over state-of-the-art unified and CFA-specific methods, indicating that adapted structural priors can enhance robust camera image restoration. The source codes and dataset are provided in the supplementary material.
}

\end{abstract}

\noindent\textbf{Code:} \url{https://github.com/Calista-pedi/SG-for-JDD.git}

\begin{figure}[!t]
\centering
\setlength{\tabcolsep}{0.7pt}
\renewcommand{\arraystretch}{1.0}
\begin{tabular}{@{}ccc@{}}
\scriptsize\bfseries ESUM &
\scriptsize\bfseries Ours &
\scriptsize\bfseries Ground Truth
\\[-0.4mm]
\includegraphics[width=0.322\columnwidth]{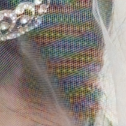} &
\includegraphics[width=0.322\columnwidth]{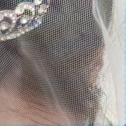} &
\includegraphics[width=0.322\columnwidth]{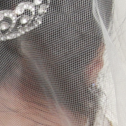}
\\[0.6mm]
\includegraphics[width=0.322\columnwidth]{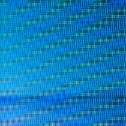} &
\includegraphics[width=0.322\columnwidth]{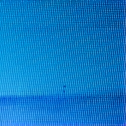} &
\includegraphics[width=0.322\columnwidth]{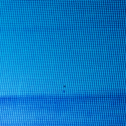}
\end{tabular}
\caption{
Visual comparison on challenging noise-free Nona-Bayer examples from MIT-Hard. Using a frozen pretrained DINOv3 encoder for structural guidance, our model better resolves ambiguous fine patterns than ESUM, suppressing false-color moir\'e and diagonal aliasing while recovering curved and periodic structures closer to the ground truth.}
\label{fig:teaser}
\end{figure}

\section{Introduction}

Recovering a clean RGB image from a noisy color filter array (CFA) observation requires jointly interpolating missing colors and suppressing sensor noise, extending the classical demosaicing problem~\cite{li2008image} to noisy sensor measurements. These two operations are closely coupled: denoising the mosaic before interpolation may remove fine structures needed for color reconstruction, whereas demosaicing noisy measurements can propagate sensor noise into spatially and cross-channel correlated artifacts. Joint demosaicing and denoising (JDD) can therefore reduce the error accumulation of independently designed sequential stages~\cite{gharbi2016deep,kokkinos2018deep,liu2020joint}. Representative approaches include the end-to-end reconstruction network of \citet{gharbi2016deep}, the cascaded residual-denoising formulation of \citet{kokkinos2018deep}, the green-channel and density-map guidance of \citet{liu2020joint}, and the joint demosaicing, denoising, and super-resolution framework of \citet{xing2021end}. Despite this progress, thin or oblique edges, repetitive textures, and moir\'e-prone patterns remain difficult to reconstruct, often resulting in false colors, zippering artifacts, distorted periodic structures, and over-smoothed details~\cite{gharbi2016deep,sharif2021sagan,tedla2025examining}.

Modern image sensors employ diverse CFA layouts, including traditional Single-Bayer and pixel-binned Quad- and Nona-Bayer patterns, whose sampling geometries and color densities differ substantially~\cite{sharif2021beyond,sharif2021sagan,lee2023efficient,tedla2025examining}. Pattern-specific networks have improved reconstruction for individual Quad- and Nona-Bayer layouts~\cite{sharif2021beyond,sharif2021sagan,zheng2024quad}, but maintaining a separate model for each layout increases deployment and model-management costs. This diversity has therefore motivated unified demosaicing and JDD methods that process multiple CFA layouts with a single network. KLAP~\cite{lee2023efficient} adapts a subset of convolutional filters according to the CFA configuration, while Tedla et al.~\cite{tedla2025examining} concatenate explicit CFA embeddings with raw measurements to jointly handle Single-, Quad-, and Nona-Bayer observations. Kumar and Yenneti~\cite{kumar2026extensible} further route CFA-specific encodings through a shared restoration core for extensible unified demosaicing. Although these methods explicitly encode how colors are sampled, their restoration mappings are still learned primarily through pixel-level reconstruction supervision. CFA masks and noise maps describe the measurement geometry and corruption level, but do not explicitly provide information about the latent edge continuation, contour organization, or periodic structure underlying an ambiguous observation. Consequently, when local CFA evidence is unreliable, particularly around oblique edges, repetitive textures, and aliasing-prone patterns, geometry-aware conditioning alone may still admit multiple plausible reconstructions. Unified JDD must therefore address not only the variation across CFA layouts, but also the structural ambiguity that remains after the sampling configuration is known, as illustrated in Fig.~\ref{fig:teaser}.

To address this limitation, we propose a structurally guided unified JDD framework that combines a pattern- and noise-conditioned restoration backbone with a frozen pretrained structural encoder that remains active during inference. First, we redesign SwinIR~\cite{liang2021swinir} for heterogeneous sensor observations. Separate shallow encoders represent the measured raw intensity, sparse color placement, and CFA/noise conditions, while deep scale-and-shift modulation injects the resulting sensor conditions into successive residual Swin Transformer blocks. This produces a sensor-aligned restoration representation that adapts to different CFA families, spatial phases, and noise levels. In parallel, a frozen DINOv3 encoder~\cite{simeoni2025dinov3} extracts dense structural representations from a sparse pseudo-RGB projection of the same observation. A lightweight convolutional adapter maps the pretrained patch representation to the spatial resolution and channel space of the restoration feature, after which a restoration-dependent residual fusion module injects the adapted representation as a structural correction rather than replacing the sensor-aligned pathway. A shared decoder trunk feeds an RGB reconstruction head and an auxiliary mosaic head, with the latter encouraging the fused representation to retain sensor-domain information. Here, structural guidance denotes the inference-time use of pretrained dense representations to provide complementary contextual cues for latent edges, contours, and repetitive structures when local CFA measurements are ambiguous.

The main contributions are summarized as follows:
\begin{itemize}
    \item We identify insufficient explicit structural guidance beyond pixel-level supervision as an important limitation of unified JDD, and introduce a structural branch that injects pretrained dense representations---instantiated with a frozen DINOv3 encoder---into CFA-aware restoration without clean-target feature supervision while remaining active at inference.

    \item We redesign SwinIR as a pattern- and noise-aware unified JDD backbone through heterogeneous sensor encoding, deep condition modulation, and native-grid feature preservation, enabling one model to adapt to different CFA families, spatial phases, and noise levels.

    \item We demonstrate consistent gains over unified and CFA-specific baselines across CFA layouts and noise levels, with particularly clear improvements in ambiguous regions such as edges, repetitive textures, and moir\'e-prone patterns, proving the effectiveness of structural guidance.
\end{itemize}


\begin{figure*}[!t]
\centering
\includegraphics[width=0.96\textwidth]{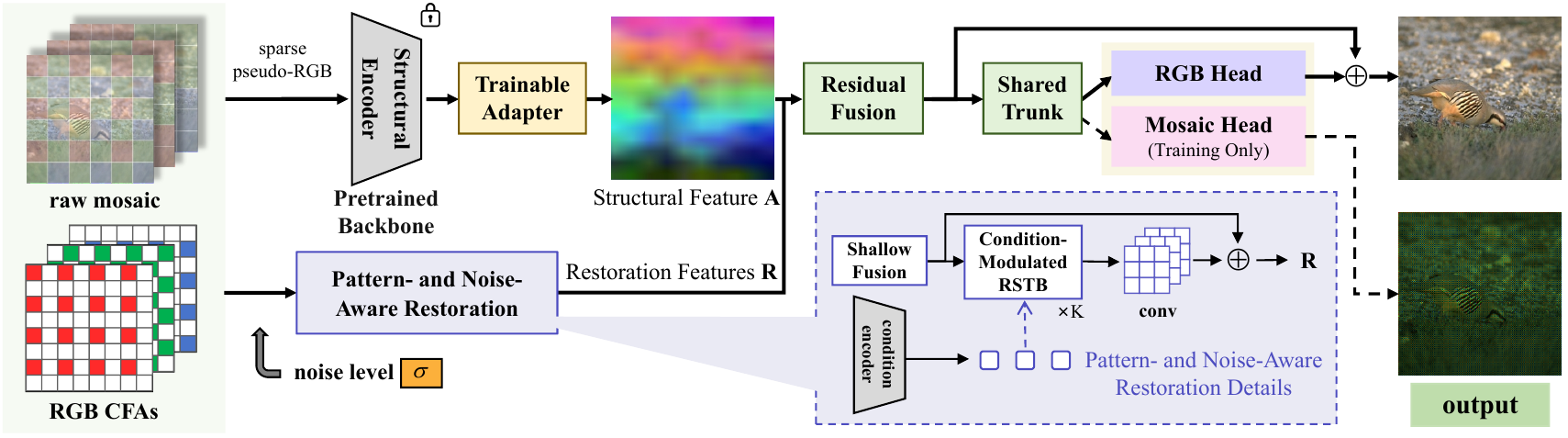}
\caption{Overview of the proposed structurally guided unified JDD framework. Unlike conventional unified restoration models that rely primarily on sensor-aligned features, our design complements the restoration feature $R$ from the pattern- and noise-aware restoration module with the pretrained structural feature $A$ through residual fusion. A shared trunk with RGB and mosaic heads supports image reconstruction and auxiliary sensor-domain supervision. The frozen structural encoder is instantiated with DINOv3 ViT-S/16 in our implementation.}
\label{fig:overview}
\end{figure*}

\section{Related Work}

\subsection{Joint and Unified CFA Restoration}

Joint demosaicing and denoising (JDD) reconstructs full-color images from noisy CFA measurements, where color interpolation and noise suppression are intrinsically coupled~\cite{bayer1976color,gunturk2005demosaicking,li2008image,zhang2011color,park2009case}. Early approaches modeled this interaction through nonparametric random fields and sequential energy minimization~\cite{khashabi2014joint,klatzer2016learning}, while later deep methods learned the reconstruction end to end using patch-based CNNs, cascaded residual denoising, and green-channel or sampling-density guidance~\cite{gharbi2016deep,kokkinos2018deep,kokkinos2019iterative,liu2020joint,ehret2019joint,guo2021joint,xing2021end}. These methods substantially improve Bayer reconstruction, but are generally designed for a fixed CFA layout.

Modern image sensors employ pixel-binned CFA patterns with different sampling geometries and color densities~\cite{sharif2021beyond,sharif2021sagan}. Pattern-specific methods address Quad-Bayer restoration through dual-encoder reconstruction or intermediate remosaicing~\cite{zheng2024quad,jia2023learning,zeng2024inheriting}, whereas unified methods share parameters across multiple sensor layouts. KLAP~\cite{lee2023efficient} adapts convolutional filters according to the CFA configuration, Tedla et al.~\cite{tedla2025examining} concatenate explicit CFA embeddings with raw measurements for Single-, Quad-, and Nona-Bayer restoration, and Kumar and Yenneti~\cite{kumar2026extensible} develop a modular architecture for multiple pixel-bin sensors. Beyond CFA adaptation through filters, masks, or embeddings, our method combines explicit CFA and noise conditioning with a pretrained structural prior for ambiguous edges and textures.

\subsection{All-in-One and Foundation-Guided Restoration}

All-in-one restoration uses degradation-aware representations, learnable prompts, or vision-language guidance~\cite{li2022airnet,potlapalli2023promptir,radford2021clip,luo2024daclip}, while MPerceiver incorporates multimodal priors~\cite{ai2024mperceiver}. Transformer backbones capture broader context through hierarchical attention, image-processing pretraining, U-shaped designs, and efficient high-resolution attention~\cite{liu2021swin,chen2021ipt,wang2022uformer,zamir2022restormer}. Dense self-supervised DINO representations~\cite{caron2021dino,oquab2024dinov2,simeoni2025dinov3} have also been adapted to low-level tasks such as nighttime dehazing~\cite{park2026siamese}.

Our method differs in its input domain and use of pretrained representations. At inference, a structural encoder extracts features from a sparse pseudo-RGB projection of noisy CFA measurements. Adapted and residually fused with sensor-aligned restoration features, they provide structural context without clean-image targets or distillation. An auxiliary mosaic head enforces sensor-domain consistency.

\section{Method}
\label{sec:method}

\subsection{Overview and Sensor Representation}
\label{subsec:overview}

Our goal is to recover a clean RGB image from noisy mosaics generated by different CFA layouts using a single model. As illustrated in Fig.~\ref{fig:overview}, the proposed framework contains two complementary pathways. A CFA- and noise-conditioned restoration branch preserves sensor-aligned observations and performs pixel-accurate reconstruction, while a frozen pretrained visual encoder provides broader structural context from a sparse pseudo-RGB representation of the same observation. The pretrained features are adapted to the restoration feature space and injected through residual fusion. A shared decoder then predicts the restored RGB image, together with an auxiliary clean mosaic used to regularize the fused representation during training.

Let $I\in[0,1]^{H\times W\times3}$ denote a clean RGB image, and let $M_p=[M_R,M_G,M_B]\in\{0,1\}^{H\times W\times3}$ denote the one-hot sampling mask of CFA configuration $p$, including its pattern family and spatial phase. The clean mosaic is formed by retaining one color measurement at each sensor location according to $M_p$, after which Gaussian noise is added to generate the noisy observation. During training, a small fraction of measurements is randomly hidden as measurement-dropout augmentation. We denote the resulting noisy mosaic and valid CFA mask by $\widetilde{Y}_{\mathrm{in}}$ and $M_p^{\mathrm{val}}$, respectively.

From the same sensor observation, we construct two input representations:
\begin{equation}
\begin{aligned}
X &= [\widetilde{Y}_{\mathrm{in}},M_p^{\mathrm{val}},S], \\
P &= M_p^{\mathrm{val}}\odot\operatorname{Rep}_3\left(\widetilde{Y}_{\mathrm{in}}\right),
\end{aligned}
\label{eq:five_channel_input}
\end{equation}
where $S$ is a normalized noise-level map and $\operatorname{Rep}_3(\cdot)$ replicates a single-channel image along the RGB channel dimension. The five-channel tensor $X\in\mathbb{R}^{H\times W\times5}$ contains the noisy mosaic, three CFA masks, and the noise map, and is used by the restoration branch. The sparse pseudo-RGB representation $P\in\mathbb{R}^{H\times W\times3}$ places each observed measurement in its corresponding color channel while leaving the two unobserved channels at zero. Unlike interpolation or remosaicing, this projection preserves the original sensor grid without introducing synthesized color values. Detailed noise synthesis and measurement-dropout settings are provided in the supplementary material.

\subsection{Pattern- and Noise-Aware Restoration}
\label{subsec:swinir}

Standard SwinIR~\cite{liang2021swinir} is designed for dense image restoration under a fixed input representation and therefore cannot directly distinguish the heterogeneous information contained in unified CFA observations. We therefore redesign SwinIR as a pattern- and noise-aware restoration backbone.

Our modification contains three coordinated components. First, a multi-source shallow encoder processes the noisy mosaic $\widetilde{Y}_{\mathrm{in}}$, the sparse pseudo-RGB observation $P$, and the sensor conditions $[M_p^{\mathrm{val}},S]$ through separate stems. The raw stem preserves measured sensor intensities, the observation stem encodes the spatial placement of the sampled colors, and the condition stem represents the CFA layout, spatial phase, and noise level. This separation prevents heterogeneous sensor information from being treated as homogeneous image channels at the network input.

\newcommand{\QualVLabel}[2][0pt]{%
    \makebox[\linewidth][c]{%
        \raisebox{#1}[0pt][0pt]{%
            \rotatebox[origin=c]{90}{%
                \scriptsize\bfseries #2%
            }%
        }%
    }%
}


\begin{figure}[!t]
\centering

\setlength{\tabcolsep}{0.45pt}
\renewcommand{\arraystretch}{1.0}

\begin{tabular}{
    @{}
    >{\centering\arraybackslash}m{0.030\columnwidth}
    *{3}{>{\centering\arraybackslash}m{0.313\columnwidth}}
    @{}
}

\QualVLabel{Reference}
&
\includegraphics[width=\linewidth]
{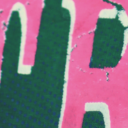}
&
\includegraphics[width=\linewidth]
{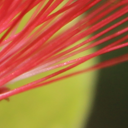}
&
\includegraphics[width=\linewidth]
{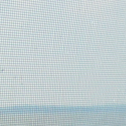}
\\[0.3mm]

\QualVLabel{w/o SG}
&
\includegraphics[width=\linewidth]
{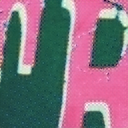}
&
\includegraphics[width=\linewidth]
{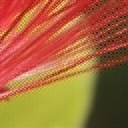}
&
\includegraphics[width=\linewidth]
{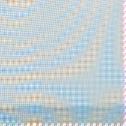}
\\[0.3mm]

\QualVLabel[0.8mm]{w SG}
&
\includegraphics[width=\linewidth]
{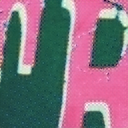}
&
\includegraphics[width=\linewidth]
{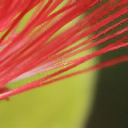}
&
\includegraphics[width=\linewidth]
{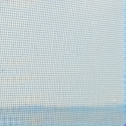}
\\[0.3mm]

\QualVLabel{Difference}
&
\includegraphics[width=\linewidth]
{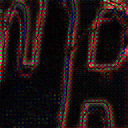}
&
\includegraphics[width=\linewidth]
{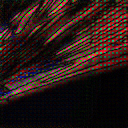}
&
\includegraphics[width=\linewidth]
{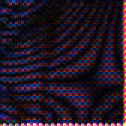}
\\[0.3mm]

\end{tabular}

\vspace{-1mm}

\caption{Qualitative comparison of our SwinIR-based model without and with pretrained structural guidance (SG), instantiated by a frozen DINOv3 encoder.}
\label{fig:dino_visual_comparison}
\end{figure}


Second, the encoded observation and sensor conditions are converted into spatial condition features $C$, which modulate both the shallow representation and successive residual Swin Transformer blocks. For the $\ell$-th restoration stage, the conditioned update is written as
\begin{equation}
\begin{aligned}
\widetilde{F}_{\ell}
&=
\left(1+\gamma_{\ell}(C)\right)\odot F_{\ell}
+
\beta_{\ell}(C), \\
F_{\ell+1}
&=
\operatorname{RSTB}_{\ell}
\left(
\widetilde{F}_{\ell}
\right),
\end{aligned}
\label{eq:deep_condition_modulation}
\end{equation}
where $\gamma_{\ell}(\cdot)$ and $\beta_{\ell}(\cdot)$ predict spatially adaptive scale and shift parameters. Unlike input-only concatenation, this deep modulation allows the shared restoration backbone to adjust its intermediate responses to different CFA families, spatial phases, and noise levels.

Third, the modified branch maintains the native $H\times W$ sensor resolution throughout restoration. This preserves the correspondence between feature locations and physical sensor measurements, which is essential for demosaicing and provides a reliable sensor-aligned representation for subsequent structural fusion. We denote the output of the modified restoration branch by
\begin{equation}
R
=
H_{\mathrm{res}}
\left(
\widetilde{Y}_{\mathrm{in}},
P,
M_p^{\mathrm{val}},
S
\right)
\in
\mathbb{R}^{H\times W\times C},
\label{eq:restoration_feature}
\end{equation}
where $C$ is the restoration feature width.

Together, the multi-source encoding, deep sensor-condition modulation, and native-grid feature preservation transform SwinIR from a conventional dense-image restorer into a unified CFA restoration backbone. The resulting feature $R$ serves as the primary reconstruction representation, while the pretrained structural branch introduced in Sec.~\ref{subsec:dinov3} provides complementary guidance for locally ambiguous structures. Detailed stem architectures and Swin Transformer configurations are provided in the supplementary material.

\begin{figure}[!t]
\centering
\includegraphics[width=\columnwidth]{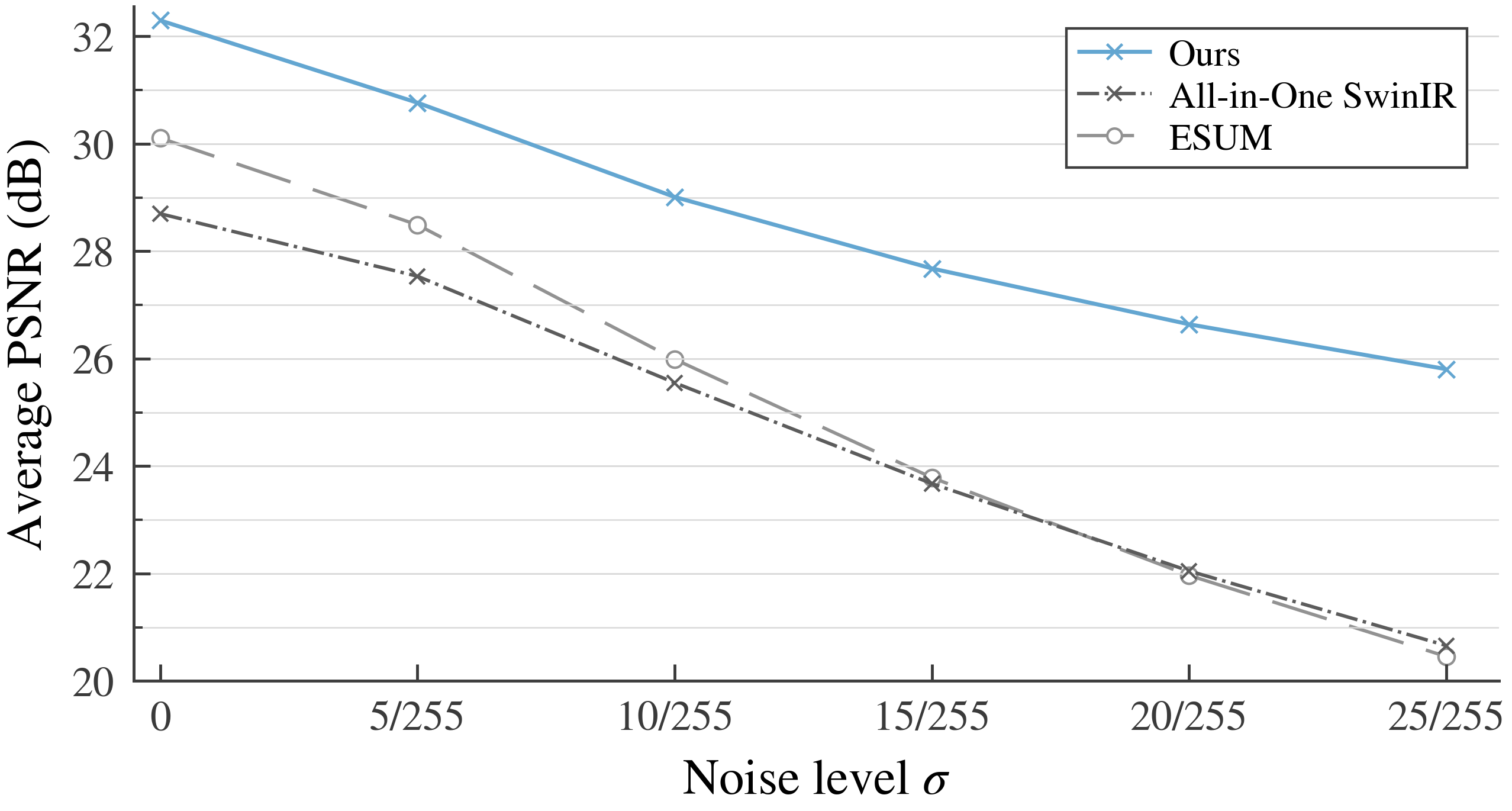}
\caption{PSNR comparison for joint demosaicing and denoising under Gaussian noise with standard deviation $\sigma$. Results are averaged over the HDR-VDP and moir\'e subsets of MIT-Hard and across Single-, Quad-, and Nona-Bayer observations.}
\label{fig:mit_hard_noise_curve}
\end{figure}

\subsection{Online Structural Prior Adaptation and Fusion}
\label{subsec:dinov3}

Let $R\in\mathbb{R}^{H\times W\times96}$ denote the sensor-aligned feature produced by the restoration branch. Directly transferring pretrained visual features is challenging because the sparse pseudo-RGB observation $P$ differs from the dense natural images used for pretraining, while the resulting patch representation is not aligned with $R$ in spatial resolution, channel width, or feature statistics. We therefore introduce an online structural pathway that converts pretrained patch tokens into restoration-compatible features and injects them as a learned residual correction.

We instantiate the frozen structural encoder $\Phi_{\mathrm{str}}$ with DINOv3 ViT-S/16~\cite{simeoni2025dinov3}. The pseudo-RGB observation $P$ is resized to $224\times224$ and normalized using the fixed pretrained input statistics. The final patch tokens are reshaped into a $14\times14$ spatial grid, bilinearly resized to $H\times W$, and projected from $384$ to $96$ channels by a lightweight adapter consisting of a $1\times1$ convolution, GELU activation, and a $3\times3$ convolution. This produces an adapted structural feature $A\in\mathbb{R}^{H\times W\times96}$.

The structural feature is integrated through restoration-dependent residual fusion:
\begin{equation}
\Delta R=f_{\mathrm{fuse}}([R,A]),
\qquad
R_f=R+\Delta R,
\label{eq:structural_fusion}
\end{equation}
where $[\cdot,\cdot]$ denotes channel-wise concatenation. The fusion module first projects the concatenated $192$-channel feature to $96$ channels using a $1\times1$ convolution, followed by GELU and a $3\times3$ convolution. Conditioning $\Delta R$ jointly on $R$ and $A$ allows the model to incorporate pretrained structural context without directly replacing the restoration representation.

The frozen structural encoder remains active during both training and inference, while the adapter, fusion module, and restoration network are jointly optimized using only the reconstruction objectives. Detailed token processing and layer configurations are provided in the supplementary material.

\subsection{Auxiliary Reconstruction and Learning Objective}
\label{subsec:decoder}

The fused representation $R_f$ is processed by a shared decoder trunk followed by lightweight RGB and mosaic heads. The RGB head includes an observation skip from $P$ and predicts the restored image $\widehat{I}\in\mathbb{R}^{H\times W\times3}$. The auxiliary head predicts the clean single-channel mosaic $\widehat{Y}\in\mathbb{R}^{H\times W}$, encouraging the shared representation to retain information associated with the original sensor domain, whereas only the RGB reconstruction is required as the task output.

We use the Charbonnier for all reconstruction terms. The full-image loss $\mathcal{L}_{\mathrm{rgb}}$ supervises the restored RGB output, $\mathcal{L}_{\mathrm{known}}$ places additional emphasis on retained sensor-color locations, and $\mathcal{L}_{\mathrm{mos}}$ supervises the auxiliary clean mosaic prediction. The overall training objective is
\begin{equation}
\mathcal{L} = \mathcal{L}_{\mathrm{rgb}} + \lambda_{\mathrm{known}}\mathcal{L}_{\mathrm{known}} + \lambda_{\mathrm{mos}}\mathcal{L}_{\mathrm{mos}},
\label{eq:final_loss}
\end{equation}
where $\lambda_{\mathrm{known}}$ and $\lambda_{\mathrm{mos}}$ control the retained-measurement and auxiliary mosaic losses, respectively.

Detailed loss definitions, network configurations, and training settings are provided in the supplementary material.

\section{Experiments}
\label{sec:experiments}

\subsection{Experimental Setup}
\label{subsec:experimental_setup}

\paragraph{Datasets.} We train on DIV2K~\cite{agustsson2017ntire}, select checkpoints on McMaster, and test on BSD100~\cite{martin2001database}, Urban100~\cite{huang2015single}, Kodak24~\cite{kodak24}, and MIT-Hard, comprising the HDR-VDP and moir\'e subsets of Gharbi et al.~\cite{gharbi2016deep}. Single-, Quad-, and Nona-Bayer observations use a common synthesis and evaluation pipeline, with one fixed validation-selected checkpoint per method and no test-specific retuning.

\paragraph{Protocol and Metrics.} We report RGB PSNR and SSIM~\cite{wang2004ssim} after cropping a two-pixel border. Unified scores are macro-averaged over the three CFA families so that Single-, Quad-, and Nona-Bayer observations contribute equally. Zero-noise experiments use $\sigma=0$; nonzero-noise results aggregate $\sigma\in\{5/255,15/255,25/255\}$, while the noise-response curve additionally includes $10/255$ and $20/255$. No checkpoint is selected or retuned on any test set.

\paragraph{Training Details.} We use $144\times144$ patches balanced across CFA families, noise up to $25/255$, and $0$--$3\%$ measurement dropout, with $\lambda_{\mathrm{known}}=0.1$ and $\lambda_{\mathrm{mos}}=0.05$. Adam~\cite{kingma2015adam}, zero weight decay, cosine annealing~\cite{loshchilov2017sgdr}, and gradient accumulation give an effective batch size of 8. The model uses $\eta_0=2\times10^{-5}$, starts from the strongest SwinIR-JDD checkpoint, and uses a DINOv3 ViT-S/16 with $224\times224$ inputs. 


\paragraph{Compared Methods.} We compare against LSUM, SRUM, and ESUM~\cite{tedla2025examining}, three unified RCAN-based JDD baselines for three CFAs. We further include two unified SwinIR-based baselines~\cite{liang2021swinir}: All-in-One SwinIR serves as the Transformer baseline, while SwinIR-JDD provides the initialization for Ours and the matched comparison in Table~\ref{tab:model_analysis}.

\subsection{Quantitative and Qualitative Results}
\label{subsec:quantitative}


\paragraph{Zero-Noise Results.}
Ours ranks first for every dataset--CFA combination in Tables~\ref{tab:zero_noise_public} and~\ref{tab:zero_noise_mit_hard}. Across BSD100, Urban100, and Kodak24, it improves the overall average over ESUM by 2.86 dB. On MIT-Hard, the gains are 2.19 dB over ESUM, and 3.60 dB over All-in-One SwinIR. The advantage remains consistent from Single- to Nona-Bayer observations, indicating that unified training does not favor one CFA family. The larger gains on Urban100 and MIT-Hard support the intended role of structural guidance on repetitive and aliasing-prone
structures.
\begin{table}[!t]
\centering

\scriptsize
\setlength{\tabcolsep}{2.2pt}
\renewcommand{\arraystretch}{1.10}

\resizebox{\columnwidth}{!}{%
\begin{tabular}{@{}lcccc@{}}
\toprule
Method & Single & Quad & Nona & Avg. \\
\midrule

LSUM / RCAN
& 27.74/.8970
& 25.96/.8251
& 25.17/.7971
& 26.29/.8397 \\

SRUM / RCAN
& 27.22/.8844
& 26.18/.8302
& 25.26/.8011
& 26.22/.8386 \\

ESUM / RCAN
& \underline{31.75/.9289}
& \underline{29.65/.9095}
& \underline{28.92/.9010}
& \underline{30.11/.9131} \\

All-in-One SwinIR
& 30.61/.9176
& 28.19/.8857
& 27.29/.8746
& 28.70/.8926 \\

\textbf{Ours}
& \textbf{33.65/.9412}
& \textbf{32.02/.9339}
& \textbf{31.22/.9299}
& \textbf{32.30/.9350} \\

\bottomrule
\end{tabular}%
}
\caption{
Zero-noise comparison of unified JDD methods on MIT-Hard across Single-, Quad-, and Nona-Bayer patterns.
}
\label{tab:zero_noise_mit_hard}
\vspace{-3mm}
\end{table}

\begin{table}[!t]
\centering
\footnotesize
\setlength{\tabcolsep}{1.5pt}
\renewcommand{\arraystretch}{1.08}
\begin{tabular}{@{}lccc@{}}
\toprule
Method & HDR-VDP & Moir\'e & All \\
  \midrule
  All-in-One SwinIR & 23.55 / .7018 & 24.36 / .6464 & 23.96 / .6741 \\
  ESUM / RCAN & 23.96 / .7092 & 24.53 / .6447 & 24.24 / .6770 \\
  \underline{SwinIR-JDD} & \underline{27.07 / .8485} & \underline{28.84 / .8265} & \underline{27.96 / .8375}\\
  \textbf{Ours} & \textbf{27.18 / .8522} & \textbf{28.99 / .8297} & \textbf{28.08 / .8409} \\
\bottomrule
\end{tabular}
\caption{
Nonzero-noise comparison of unified methods on MIT-Hard, averaged across Single-, Quad-, and Nona-Bayer observations at $\sigma\in\{5/255,15/255,25/255\}$.
}
\label{tab:mit_hard_nonzero}
\vspace{-3mm}
\end{table}

\begin{figure*}[!t]
\centering

\newlength{\UrbanImageH}
\setlength{\UrbanImageH}{0.142\textwidth}

\newcommand{\UrbanImage}[1]{%
    \includegraphics[height=\UrbanImageH]{#1}%
}

\newcommand{\UrbanRowLabel}[1]{%
    \parbox[b][\UrbanImageH][c]{3.2mm}{%
        \centering
        \rotatebox[origin=c]{90}{\scriptsize\bfseries #1}%
    }%
}

\setlength{\tabcolsep}{0.4pt}
\renewcommand{\arraystretch}{1.0}

\begin{tabular}{@{}c*{6}{c}@{}}

&
\scriptsize\bfseries full image
&
\scriptsize\bfseries LSUM
&
\scriptsize\bfseries ESUM
&
\scriptsize\bfseries All-in-One SwinIR
&
\scriptsize\bfseries Ours
&
\scriptsize\bfseries GT
\\[-0.4mm]

\UrbanRowLabel{Single-Bayer}
&
\includegraphics[
    width=0.215\textwidth,
    height=\UrbanImageH
]{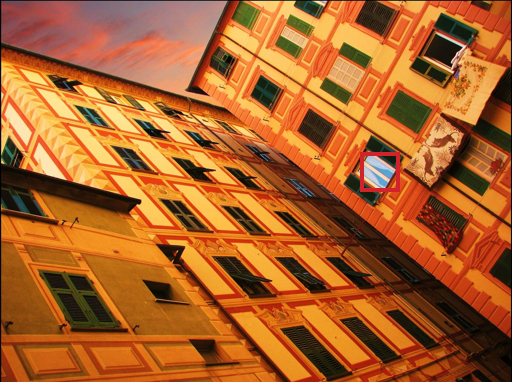}
&
\UrbanImage{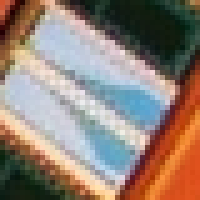}
&
\UrbanImage{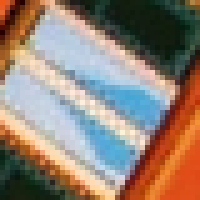}
&
\UrbanImage{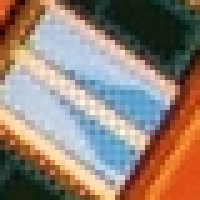}
&
\UrbanImage{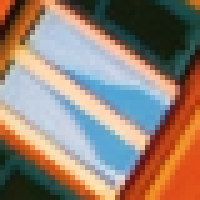}
&
\UrbanImage{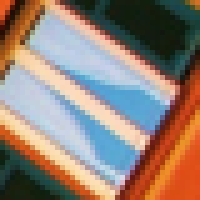}
\\[0.3mm]

\UrbanRowLabel{Quad-Bayer}
&
\UrbanImage{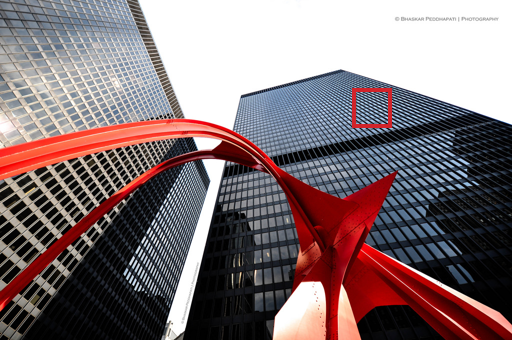}
&
\UrbanImage{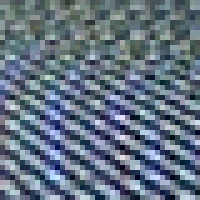}
&
\UrbanImage{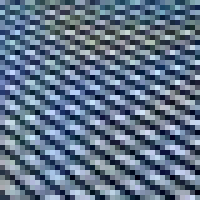}
&
\UrbanImage{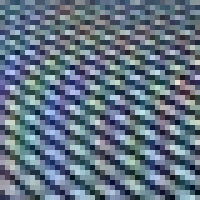}
&
\UrbanImage{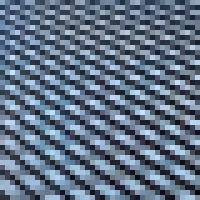}
&
\UrbanImage{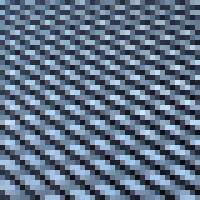}
\\[0.3mm]

\UrbanRowLabel{Nona-Bayer}
&
\UrbanImage{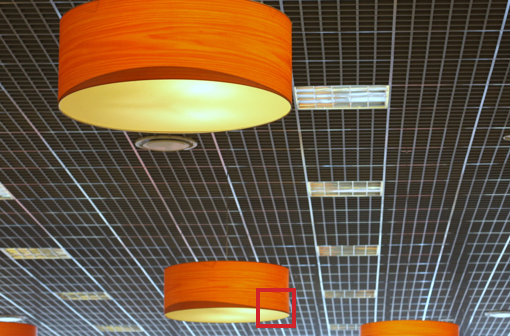}
&
\UrbanImage{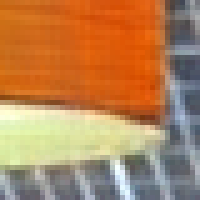}
&
\UrbanImage{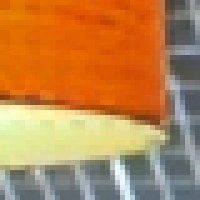}
&
\UrbanImage{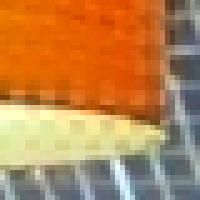}
&
\UrbanImage{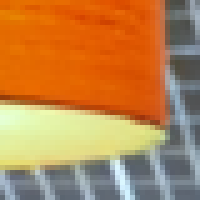}
&
\UrbanImage{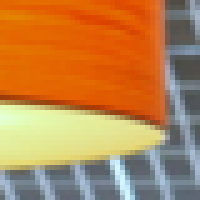}

\end{tabular}

\caption{Comparison on Urban100 at $\sigma=0$. LSUM and ESUM~\cite{tedla2025examining} are unified JDD baselines, while All-in-One SwinIR is our matched SwinIR-based baseline~\cite{liang2021swinir}. Competing methods show false-color moir\'e, zippering, and distorted repetitive patterns, especially under sparser Quad- and Nona-Bayer observations.}
\label{fig:urban100_methods}
\end{figure*}

\begin{figure*}[!t]
\centering
\setlength{\tabcolsep}{0.45pt}
\renewcommand{\arraystretch}{1.0}

\begin{tabular}{@{}c*{6}{c}@{}}


&
\scriptsize\bfseries Reference
&
\scriptsize\bfseries $\sigma=0$
&
\scriptsize\bfseries $\sigma=15/255$
&
\scriptsize\bfseries $\Delta$ vs.\ $\sigma=0$
&
\scriptsize\bfseries $\sigma=25/255$
&
\scriptsize\bfseries $\Delta$ vs.\ $\sigma=0$
\\[-0.4mm]

\parbox[b][0.158\textwidth][c]{0.022\textwidth}{
    \centering
    \rotatebox[origin=c]{90}{\scriptsize\bfseries Scene 1}
}
&
\FigureOrPlaceholder
    {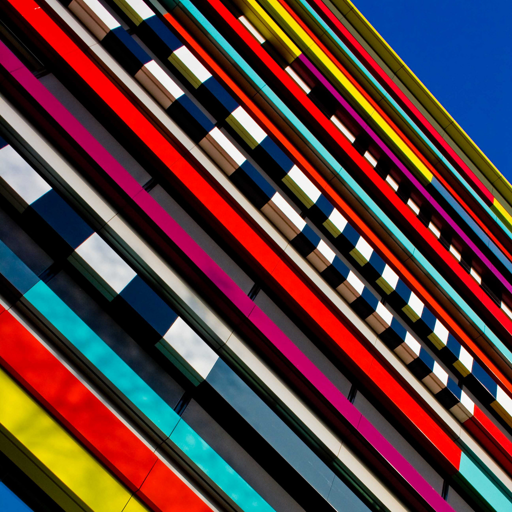}
    {0.158\textwidth}
    {0.158\textwidth}
    {Scene 1 ground truth}
&
\FigureOrPlaceholder
    {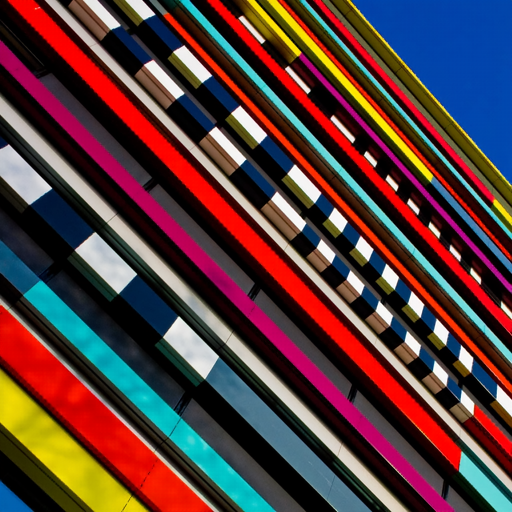}
    {0.158\textwidth}
    {0.158\textwidth}
    {Scene 1 restored by Ours at zero noise}
&
\FigureOrPlaceholder
    {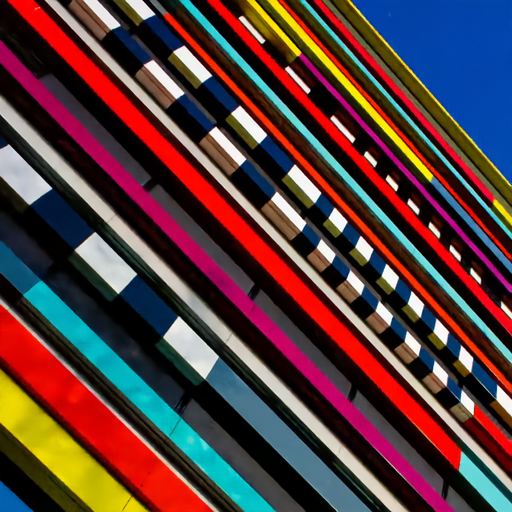}
    {0.158\textwidth}
    {0.158\textwidth}
    {Scene 1 restored by Ours at $15/255$}
&
\FigureOrPlaceholder
    {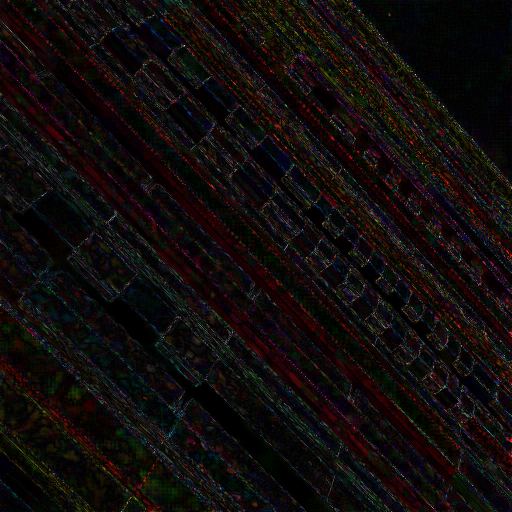}
    {0.158\textwidth}
    {0.158\textwidth}
    {Absolute difference between the outputs at $15/255$ and zero noise}
&
\FigureOrPlaceholder
    {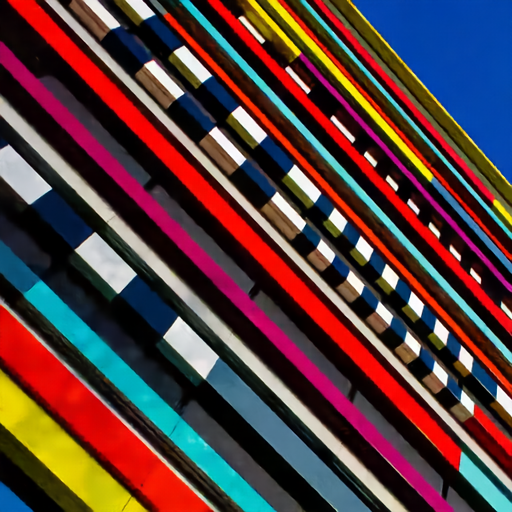}
    {0.158\textwidth}
    {0.158\textwidth}
    {Scene 1 restored by Ours at $25/255$}
&
\FigureOrPlaceholder
    {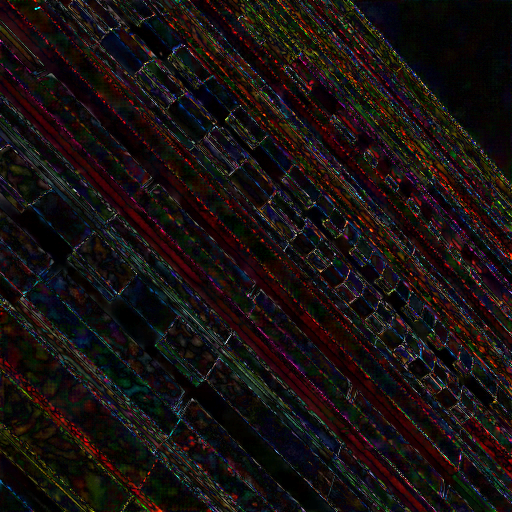}
    {0.158\textwidth}
    {0.158\textwidth}
    {Absolute difference between the outputs at $25/255$ and zero noise}
\\[0.6mm]

\parbox[b][0.158\textwidth][c]{0.022\textwidth}{
    \centering
    \rotatebox[origin=c]{90}{\scriptsize\bfseries Scene 2}
}
&
\FigureOrPlaceholder
    {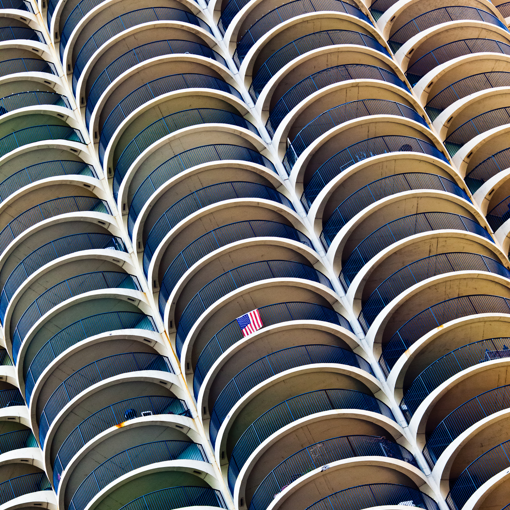}
    {0.158\textwidth}
    {0.158\textwidth}
    {Scene 2 ground truth}
&
\FigureOrPlaceholder
    {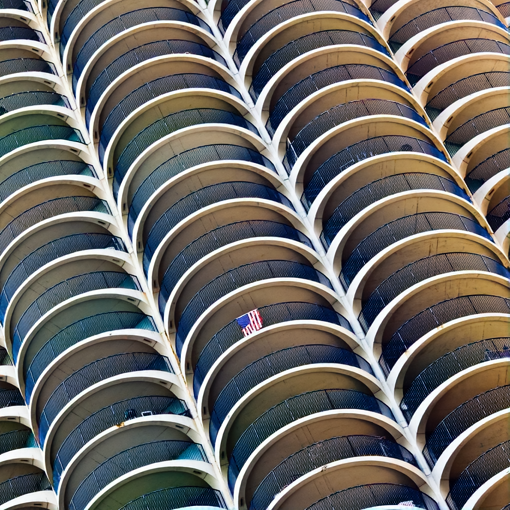}
    {0.158\textwidth}
    {0.158\textwidth}
    {Scene 2 restored by Ours at zero noise}
&
\FigureOrPlaceholder
    {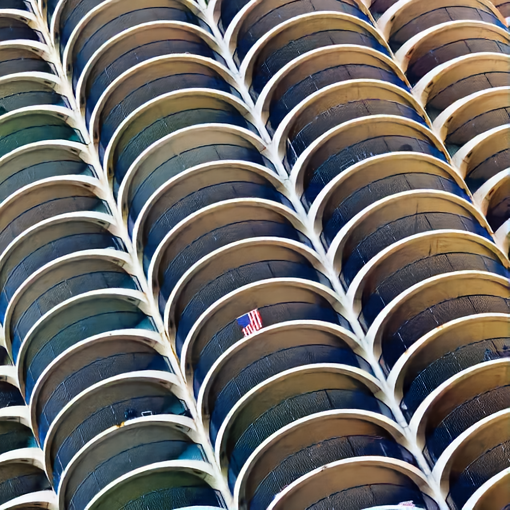}
    {0.158\textwidth}
    {0.158\textwidth}
    {Scene 2 restored by Ours at $15/255$}
&
\FigureOrPlaceholder
    {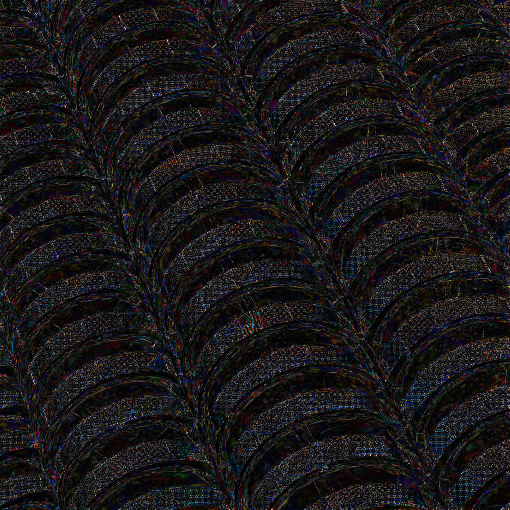}
    {0.158\textwidth}
    {0.158\textwidth}
    {Absolute difference between the outputs at $15/255$ and zero noise}
&
\FigureOrPlaceholder
    {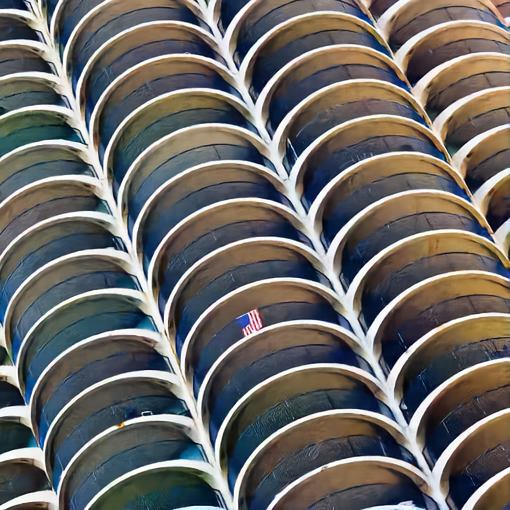}
    {0.158\textwidth}
    {0.158\textwidth}
    {Scene 2 restored by Ours at $25/255$}
&
\FigureOrPlaceholder
    {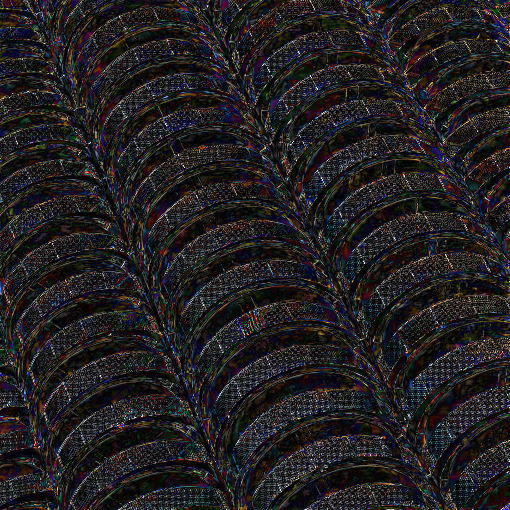}
    {0.158\textwidth}
    {0.158\textwidth}
    {Absolute difference between the outputs at $25/255$ and zero noise}

\end{tabular}
\caption{
Visual comparison of our model under different noise levels.Scenes 1 and 2 respectively use Quad- and Nona-Bayer observations.
}
\label{fig:ours_noise_recovery}
\end{figure*}

\begin{table*}[!t]
\centering

\scriptsize
\setlength{\tabcolsep}{1.6pt}
\renewcommand{\arraystretch}{1.10}

\resizebox{\textwidth}{!}{%
\begin{tabular}{@{}l*{10}{c}@{}}
\toprule

\multirow{2}{*}{Method}
& \multicolumn{3}{c}{BSD100}
& \multicolumn{3}{c}{Urban100}
& \multicolumn{3}{c}{Kodak24}
& \multirow{2}{*}{\shortstack{Overall\\Avg.}} \\

\cmidrule(lr){2-4}
\cmidrule(lr){5-7}
\cmidrule(lr){8-10}

& Single & Quad & Nona
& Single & Quad & Nona
& Single & Quad & Nona
& \\

\midrule

LSUM / RCAN
& 29.35/.9575 & 27.86/.9106 & 27.28/.8852
& 25.26/.9252 & 24.38/.8835 & 24.05/.8571
& 32.36/.9730 & 30.03/.9270 & 29.28/.9038
& 27.76/.9137 \\

SRUM / RCAN
& 27.27/.9582 & 28.25/.9157 & 27.52/.8893
& 24.81/.9258 & 24.39/.8838 & 23.53/.8547
& 29.68/.9697 & 30.35/.9310 & 29.60/.9096
& 27.27/.9153 \\

ESUM / RCAN
& \underline{38.17/.9823}
& \underline{35.08/.9696}
& \underline{34.85/.9679}
& \underline{34.45/.9703}
& \underline{32.15/.9575}
& \underline{31.66/.9537}
& \underline{39.67/.9825}
& \underline{36.00/.9680}
& \underline{36.06/.9680}
& \underline{35.34/.9689} \\

All-in-One SwinIR
& 36.79/.9799 & 33.45/.9628 & 33.16/.9611
& 31.99/.9600 & 29.10/.9343 & 28.90/.9303
& 37.81/.9793 & 33.54/.9572 & 33.75/.9581
& 33.17/.9581 \\

\textbf{Ours}
& \textbf{40.94/.9903}
& \textbf{38.19/.9842}
& \textbf{37.82/.9834}
& \textbf{37.32/.9812}
& \textbf{35.48/.9758}
& \textbf{34.68/.9733}
& \textbf{41.74/.9879}
& \textbf{39.00/.9816}
& \textbf{38.65/.9810}
& \textbf{38.20/.9821} \\

\bottomrule
\end{tabular}%
}
\caption{Zero-noise comparison on BSD100, Urban100, and Kodak24. Avg.\ denotes the unweighted mean over all nine dataset-pattern combinations. Best and second-best results are respectively shown in bold and underlined.}
\label{tab:zero_noise_public}
\end{table*}

\begin{figure}[!t]
\centering

\setlength{\tabcolsep}{0.45pt}
\renewcommand{\arraystretch}{1.0}

\begin{tabular}{
    @{}
    >{\centering\arraybackslash}m{0.030\columnwidth}
    *{3}{>{\centering\arraybackslash}m{0.313\columnwidth}}
    @{}
}

&
{\scriptsize\bfseries Single-Bayer}
&
{\scriptsize\bfseries Quad-Bayer}
&
{\scriptsize\bfseries Nona-Bayer}
\\[-0.3mm]

\QualVLabel{LSUM}
&
\includegraphics[width=\linewidth]
{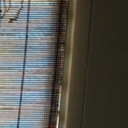}
&
\includegraphics[width=\linewidth]
{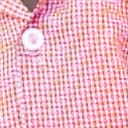}
&
\includegraphics[width=\linewidth]
{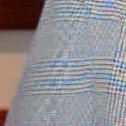}
\\[0.3mm]

\QualVLabel{ESUM}
&
\includegraphics[width=\linewidth]
{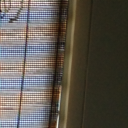}
&
\includegraphics[width=\linewidth]
{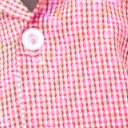}
&
\includegraphics[width=\linewidth]
{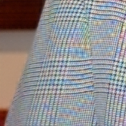}
\\[0.3mm]

\QualVLabel[0.8mm]{All-in-One SwinIR}
&
\includegraphics[width=\linewidth]
{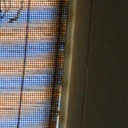}
&
\includegraphics[width=\linewidth]
{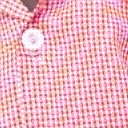}
&
\includegraphics[width=\linewidth]
{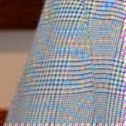}
\\[0.3mm]

\QualVLabel{Ours}
&
\includegraphics[width=\linewidth]
{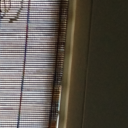}
&
\includegraphics[width=\linewidth]
{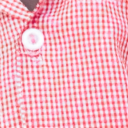}
&
\includegraphics[width=\linewidth]
{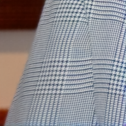}
\\[0.3mm]

\QualVLabel[0.5mm]{Ground Truth}
&
\includegraphics[width=\linewidth]
{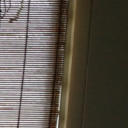}
&
\includegraphics[width=\linewidth]
{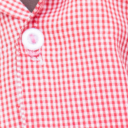}
&
\includegraphics[width=\linewidth]
{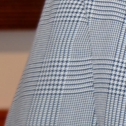}

\end{tabular}

\vspace{-1mm}

\caption{
Comparison on MIT-Hard~\cite{gharbi2016deep} at $\sigma=0$ across Single-, Quad-, and Nona-Bayer observations.}
\label{fig:qualitative}
\end{figure}

\paragraph{Qualitative Results.}
Figure~\ref{fig:qualitative} complements the dataset averages with difficult MIT-Hard crops. The baselines often introduce false-color moir\'e, zippering, or distorted periodic patterns, whereas our method better preserves line spacing, edge continuity, and color organization. The clearest improvements occur where sparse local measurements remain ambiguous but the underlying contour or repetition is recognizable. Figs.~\ref{fig:dino_visual_comparison} and~\ref{fig:urban100_methods} show similar changes on thin strokes, repeated grids, and fabric-like textures. All methods use identical observations, crop coordinates, and display ranges; the difference maps in Fig.~\ref{fig:dino_visual_comparison} visualize output changes rather than errors against the ground truth.



\begin{table}[!t]
\centering
\footnotesize
\setlength{\tabcolsep}{1.5pt}
\renewcommand{\arraystretch}{1.08}
\begin{tabular}{@{}lccc@{}}
  \toprule
  Variant & Bal. & $\sigma=0$ & $\sigma=25/255$ \\
  \midrule
  Dual-head
  & 32.461 & 35.637 & 29.087 \\

  $+$ Frozen structural encoder
  & 32.477 & 35.651 & 29.101 \\

  $+$ Shared trunk
  & 32.505 & 35.696 & 29.114 \\

  $+$ Fine-tuning
  & 32.525 & 35.726 & 29.137 \\

  \textbf{Full model}
  & \textbf{32.732} & \textbf{36.096} & \textbf{29.249} \\

  \bottomrule
  \end{tabular}

\caption{Model-development analysis in PSNR. Sequential variants trace the progression from the dual-head baseline to structural guidance, shared decoding, and fine-tuning, while the Full model achieves the best performance.}
\label{tab:model_analysis}
\end{table}


\paragraph{Noise Robustness.}
Table~\ref{tab:mit_hard_nonzero} and Figs.~\ref{fig:mit_hard_noise_curve} and~\ref{fig:ours_noise_recovery} evaluate robustness under synthetic
sensor noise. Averaged over three CFAs and three nonzero noise levels, Our method outperforms ESUM by 3.84 dB and All-in-One SwinIR by 4.13 dB. The gains appear on both HDR-VDP and moir\'e, rather than being driven by one subset. The noise-response curve and visual examples further show that the advantage persists as corruption increases without removing dominant boundaries and repetitive structures.


\subsection{Developmental and Matched-Start Analysis}
\label{subsec:analysis}

Table~\ref{tab:model_analysis} traces the development from the dual-head baseline to the complete structurally guided model. Adding the frozen structural encoder, shared trunk, and subsequent fine-tuning progressively raises the balanced PSNR from 32.461 to 32.525~dB. The final configuration integrates these components and achieves the best balanced PSNR of 32.732~dB.

\paragraph{Model Analysis.} Table~\ref{tab:model_analysis} separates the contribution of structural guidance from the larger gains provided by the noise-aware restoration backbone. Relative to SwinIR-JDD, the final model gains 0.12 dB and 0.0034 SSIM on MIT-Hard. Because the earlier sequential variants use different initializations, they describe model development rather than a strictly additive ablation.

Early dual-head, DINOv3-branch, and shared-decoder variants were initialized sequentially from the earlier checkpoint, whereas the final model was rebuilt from SwinIR-JDD. The early sequence is therefore reported as a \emph{developmental analysis}, not as a strict one-factor additive ablation. The matched-start comparison between SwinIR-JDD and the complete model initialized from it measures the complete architectural extension rather than isolating DINOv3 alone. The early series suggests that the decoder and structural branch can be optimized successfully, but the small increments cannot be separated from checkpoint quality and training stage. The full-model result shows that the strongest performance is obtained when the complete architecture starts from an already reliable pixel-level JDD mapping. We therefore attribute the large difference from external baselines to the complete system and training recipe, while treating DINOv3 as a complementary structural component rather than the sole source of the gain.

\section{Conclusion and Limitations}

We presented a structural-reasoning-guided framework for unified JDD across Single-, Quad-, and Nona-Bayer patterns and multiple noise levels. A CFA- and noise-conditioned SwinIR pathway provides spatially aligned restoration features, while a frozen pretrained structural encoder supplies complementary dense representations from the sparse sensor observation. Feature adaptation, residual fusion, and auxiliary mosaic reconstruction integrate these structural cues while retaining sensor-domain information. The resulting system achieves consistent improvements across standard benchmarks and structurally challenging scenes under both zero and nonzero noise. Current limitations include the use of synthetic Gaussian noise and sRGB-derived mosaics, the absence of calibrated real multi-CFA RAW data, and the inference cost of the structural pathway. The architecture accepts alternative pretrained encoders through its adapter interface, although the experiments instantiate only DINOv3. Future work will evaluate other structural encoders, realistic signal-dependent noise, and more efficient structural guidance.

\bibliography{references}

\appendix

\clearpage

\twocolumn[
\begin{center}
    {\LARGE\bfseries Appendix}
    \vspace{1em}
\end{center}
]

\section{Supplementary Overview}
\label{sec:supp_overview}

This supplementary document complements the main paper with implementation details and expanded evidence rather than repeating its central narrative. Section~\ref{sec:supp_method} specifies the observation synthesis, conditioned restoration stream, online structural reasoning branch, shared dual-head decoder, and training objectives. Section~\ref{sec:supp_protocol} documents the datasets, staged optimization, model selection, metric implementation, and comparison scope. Section~\ref{sec:supp_results} provides expanded MIT-Hard results, model-development analysis, paired validation diagnostics, the implementation distinction from DINOv2 distillation, and computational cost. Sections~\ref{sec:supp_reproducibility} and~\ref{sec:supp_limitations} state the reproducibility boundary, limitations, and ethical considerations.

The experimental evidence is organized around four questions: whether the unified architecture behaves consistently across Single-, Quad-, and Nona-Bayer observations; whether performance degrades predictably as noise and CFA block size increase; whether the reported validation gains persist across individual cases rather than only in aggregate averages; and what computation is required by the online structural branch. Aggregate, group-wise, and paired results answer complementary parts of these questions and are interpreted only within their stated comparison protocols.

\section{Additional Method Details}
\label{sec:supp_method}

\paragraph{Notation.} We omit the batch dimension throughout. Bold symbols denote images, masks, vectors, or feature tensors; $H$ and $W$ denote the original spatial height and width, while $H_p=8\lceil H/8\rceil$ and $W_p=8\lceil W/8\rceil$ denote the reflect-padded dimensions used internally by the network. The index $c\in\{R,G,B\}$ denotes color channels, $p\in\mathcal{P}=\{\mathrm{Single},\mathrm{Quad},\mathrm{Nona}\}$ denotes CFA families, and $u$ denotes spatial locations. The operators $\odot$, $[\cdot,\ldots,\cdot]$, $\operatorname{Rep}_{3}$, and $\operatorname{mean}$ denote elementwise multiplication, channelwise concatenation, threefold channel replication, and averaging over the indicated indices, respectively. A tilde marks a corrupted observation, a hat marks a prediction, and the superscript ``$\mathrm{val}$'' marks measurements remaining valid after dropout. Symbols $h(\cdot)$, $f(\cdot)$, and $H(\cdot)$ denote learned network modules. Scalar maps are broadcast only where stated explicitly; the scale and shift used by condition modulation are instead predicted for every spatial token and feature channel.

\subsection{Unified Observation Synthesis}
\label{sec:supp_observation}

\paragraph{CFA construction.} Let $\mathbf{I}\in[0,1]^{3\times H\times W}$ be a clean RGB image, and let $\mathbf{M}^{p}_{c}\in\{0,1\}^{H\times W}$ be its binary sampling mask for channel $c$ under CFA family $p$. Exactly one color is sampled at every sensor location before measurement dropout, so $\sum_c\mathbf{M}^{p}_{c}=\mathbf{1}$. The three CFA families use same-color block sizes $k\in\{1,2,3\}$ and therefore have spatial periods $2$, $4$, and $6$, respectively. Although the mask generator supports the four standard Bayer phases, all reported training, validation, and test observations use the fixed RGGB phase for each block size; CFA phase is not randomly sampled during training.

\paragraph{Noise synthesis.} The clean sensor-domain mosaic is
\begin{equation}
\mathbf{Y}=\sum_{c\in\{R,G,B\}}\mathbf{M}^{p}_{c}\odot\mathbf{I}_{c}.
\label{eq:supp_clean_mosaic}
\end{equation}
For each training crop, the noise standard deviation is sampled as $\sigma\sim\mathcal{U}(0,0.098)$, whose upper bound is approximately $25/255$. Signal-independent Gaussian noise is added before clipping:
\begin{equation}
\widetilde{\mathbf{Y}}=\operatorname{clip}(\mathbf{Y}+\mathbf{n},0,1),\qquad n_u\overset{\mathrm{i.i.d.}}{\sim}\mathcal{N}(0,\sigma^{2}).
\label{eq:supp_noise}
\end{equation}
Here, $n_u$ is the noise at sensor location $u$, ``i.i.d.'' indicates that these samples are independent and identically distributed, and $\operatorname{clip}(\cdot,0,1)$ truncates the noisy mosaic to the valid intensity range.

\paragraph{Measurement dropout.} Measurement dropout is used only during training. A dropout probability $q\sim\mathcal{U}(0,0.05)$ is sampled independently for each crop, after which each spatial entry of the retention mask satisfies $\mathbf{B}_{u}\sim\operatorname{Bernoulli}(1-q)$. The input mosaic and valid CFA masks are
\begin{equation}
\widetilde{\mathbf{Y}}_{\mathrm{in}}=\mathbf{B}\odot\widetilde{\mathbf{Y}},\qquad \mathbf{M}^{p,\mathrm{val}}_{c}=\mathbf{B}\odot\mathbf{M}^{p}_{c}.
\label{eq:supp_dropout}
\end{equation}
The clean retained-mosaic target used by the auxiliary head follows the same validity mask:
\begin{equation}
\mathbf{Y}^{\mathrm{val}}=\sum_{c\in\{R,G,B\}}\mathbf{M}^{p,\mathrm{val}}_{c}\odot\mathbf{I}_{c}=\mathbf{B}\odot\mathbf{Y}.
\label{eq:supp_retained_mosaic}
\end{equation}
The normalized noise map is
\begin{equation}
\mathbf{S}=\mathbf{B}\odot\operatorname{clip}\left(\frac{\sigma}{0.1},0,1\right).
\label{eq:supp_sigma_map}
\end{equation}
Here, $\mathbf{B}_{u}=1$ and $0$ denote a retained and dropped measurement, respectively, and the scalar normalized noise level in Eq.~\eqref{eq:supp_sigma_map} is broadcast across the sensor grid before multiplication by $\mathbf{B}$. Consequently, a dropped location is marked unavailable in the mosaic, all three CFA masks, and the noise map, whereas a physically valid zero-valued measurement remains distinguishable because its corresponding CFA mask equals one. Standard evaluation uses $\mathbf{B}=\mathbf{1}$.

\paragraph{Network representations.} The unified five-channel input to the restoration stream is
\begin{equation}
\mathbf{X}=\left[\widetilde{\mathbf{Y}}_{\mathrm{in}},\mathbf{M}^{p,\mathrm{val}}_{R},\mathbf{M}^{p,\mathrm{val}}_{G},\mathbf{M}^{p,\mathrm{val}}_{B},\mathbf{S}\right]\in\mathbb{R}^{5\times H\times W}.
\label{eq:supp_five_channel}
\end{equation}
Let $\mathbf{M}^{p,\mathrm{val}}=[\mathbf{M}^{p,\mathrm{val}}_{R},\mathbf{M}^{p,\mathrm{val}}_{G},\mathbf{M}^{p,\mathrm{val}}_{B}]$. The sparse pseudo-RGB projection is
\begin{equation}
\mathbf{I}_{\mathrm{obs}}=\mathbf{M}^{p,\mathrm{val}}\odot\operatorname{Rep}_{3}(\widetilde{\mathbf{Y}}_{\mathrm{in}})\in\mathbb{R}^{3\times H\times W}.
\label{eq:supp_sparse_rgb}
\end{equation}
This projection places each retained sensor measurement in its observed color channel and sets the two unavailable color channels to zero; it introduces no interpolated color values and uses no clean RGB information.

\subsection{Pattern- and Noise-Aware Restoration Stream}
\label{sec:supp_restoration}

\paragraph{Sensor-aligned stems.} The restoration stream forms three native-resolution feature maps:
\begin{align}
\mathbf{F}_{\mathrm{raw}}&=h_{\mathrm{raw}}(\widetilde{\mathbf{Y}}_{\mathrm{in}}),\\
\mathbf{F}_{\mathrm{obs}}&=h_{\mathrm{obs}}(\mathbf{I}_{\mathrm{obs}}),\\
\mathbf{F}_{\mathrm{cond}}&=h_{\mathrm{cond}}([\mathbf{M}^{p,\mathrm{val}}_{R},\mathbf{M}^{p,\mathrm{val}}_{G},\mathbf{M}^{p,\mathrm{val}}_{B},\mathbf{S}]).
\label{eq:supp_stem_features}
\end{align}
The features $\mathbf{F}_{\mathrm{raw}}$, $\mathbf{F}_{\mathrm{obs}}$, and $\mathbf{F}_{\mathrm{cond}}$ encode the raw mosaic intensity, its color-aligned sparse projection, and the CFA--noise condition, respectively. All three stems preserve spatial resolution and produce 96 channels. The raw stem is a single $3\times3$ convolution. The observation and condition stems each use a $3\times3$ convolution, GELU, and a second $3\times3$ convolution. Their outputs are concatenated into a 288-channel tensor and compressed to 96 channels by $h_{\mathrm{fuse}}$, which uses a $1\times1$ convolution, GELU, and a $3\times3$ refinement convolution:
\begin{equation}
\mathbf{F}_{0}=\mathbf{F}_{\mathrm{raw}}+h_{\mathrm{fuse}}([\mathbf{F}_{\mathrm{raw}},\mathbf{F}_{\mathrm{obs}},\mathbf{F}_{\mathrm{cond}}]).
\label{eq:supp_initial_feature}
\end{equation}
The fused initial feature $\mathbf{F}_{0}$ therefore retains a direct residual path from $\mathbf{F}_{\mathrm{raw}}$. All convolutions in these mappings use unit stride and a learnable bias. The $3\times3$ convolutions use one-pixel zero padding, while the $1\times1$ projection uses no padding. No normalization layer is used inside the stems or $h_{\mathrm{fuse}}$; LayerNorm is applied subsequently by the patch-embedding and transformer stages.

\begin{table}[!t]
\centering
\footnotesize
\setlength{\tabcolsep}{8.0pt}
\renewcommand{\arraystretch}{1.08}
\begin{tabular}{@{}llll@{}}
\toprule
Mapping & Input & Output & Operation \\
\midrule
$h_{\mathrm{raw}}$ & 1 & 96 & $3\times3$ conv \\
$h_{\mathrm{obs}}$ & 3 & 96 & $3\times3$ conv, GELU, $3\times3$ conv \\
$h_{\mathrm{cond}}$ & 4 & 96 & $3\times3$ conv, GELU, $3\times3$ conv \\
$h_{\mathrm{fuse}}$ & 288 & 96 & $1\times1$ conv, GELU, $3\times3$ conv \\
\bottomrule
\end{tabular}
\caption{Layer configurations of the restoration stems and their fusion mapping. All operations preserve the sensor-grid resolution and use no internal normalization.}
\label{tab:supp_stems}
\end{table}

\paragraph{Conditioned Swin Transformer.} The condition representation combines the CFA--noise feature with the observation-aligned feature before patch embedding:
\begin{equation}
\mathbf{C}=\operatorname{PatchEmbed}(\mathbf{F}_{\mathrm{cond}}+\mathbf{F}_{\mathrm{obs}})\in\mathbb{R}^{H_pW_p\times96}.
\label{eq:supp_condition_tokens}
\end{equation}
The same spatially aligned condition tokens are injected before the transformer body and before each residual Swin Transformer block (RSTB). A token-wise two-layer modulation MLP applies
\begin{equation}
\operatorname{CM}(\mathbf{F};\mathbf{C})=\mathbf{F}\odot\left(1+0.1\tanh(\boldsymbol{\gamma}_{\mathbf{C}})\right)+\boldsymbol{\beta}_{\mathbf{C}}.
\label{eq:supp_condition_modulation}
\end{equation}
Here, $\boldsymbol{\gamma}_{\mathbf{C}},\boldsymbol{\beta}_{\mathbf{C}}\in\mathbb{R}^{H_pW_p\times96}$ are scale and shift tensors predicted for each spatial token and channel, not channelwise constants broadcast across the image. The factor $0.1$ constrains the multiplicative scale to $(0.9,1.1)$; because the final modulation layer is zero-initialized, the conditioning path begins as the identity. The sensor-aware restoration feature is
\begin{equation}
\mathbf{R}=\mathbf{F}_{0}+H_{\mathrm{conv}}\!\left(H_{\mathrm{RSTB}}(\mathbf{F}_{0};\mathbf{C})\right).
\label{eq:supp_restoration_feature}
\end{equation}
Here, $H_{\mathrm{RSTB}}$ denotes the conditioned transformer body, the semicolon indicates modulation by $\mathbf{C}$ rather than channel concatenation, and $H_{\mathrm{conv}}$ is the post-body convolutional mapping. The resulting $\mathbf{R}$ is the sensor-aware restoration feature. The body uses four RSTBs, six Swin Transformer layers per RSTB, an embedding dimension of 96, six attention heads, window size eight, MLP ratio two, patch size one, and a single $3\times3$ convolution for each RSTB residual connection. The drop-path rate is 0.1, attention and feature dropout are zero, QKV bias and patch normalization are enabled, absolute positional embedding is disabled, and gradient recomputation is enabled during training. Inputs are reflect-padded to a multiple of the window size before either branch is evaluated; the structural branch therefore receives the padded pseudo-RGB observation, and the final RGB and mosaic predictions are cropped back to the original spatial extent.

\begin{table}[!t]
\centering
\footnotesize
\setlength{\tabcolsep}{4.0pt}
\renewcommand{\arraystretch}{1.08}
\begin{tabular}{@{}ll@{}}
\toprule
Configuration & Value \\
\midrule
Feature dimension & 96 \\
Patch size & 1 \\
RSTBs / layers per RSTB & 4 / 6 \\
Attention heads & 6 \\
Window size & 8 \\
MLP ratio & 2 \\
RSTB residual mapping & one $3\times3$ convolution \\
\bottomrule
\end{tabular}
\caption{Core configuration of the conditioned SwinIR restoration stream~\cite{liang2021swinir}.}
\label{tab:supp_restoration_config}
\end{table}

\subsection{Online Structural Reasoning Branch}
\label{sec:supp_structural}

\paragraph{Residual pseudo-RGB stem.} In addition to the fixed sparse projection in Eq.~\eqref{eq:supp_sparse_rgb}, the implementation defines a residual pseudo-RGB stem $g_{\mathrm{stem}}$ with a $3\times3$ convolution from 5 to 48 channels, GELU, and a second $3\times3$ convolution from 48 to 3 channels:
\begin{equation}
\mathbf{P}=\operatorname{clip}\!\left(\mathbf{I}_{\mathrm{obs}}+0.1\tanh(g_{\mathrm{stem}}(\mathbf{X})),0,1\right).
\label{eq:supp_pseudo_rgb_stem}
\end{equation}
Here, $g_{\mathrm{stem}}(\mathbf{X})$ predicts a three-channel correction; the scaled hyperbolic tangent bounds it to $(-0.1,0.1)$ before clipping, and $\mathbf{P}$ is the pseudo-RGB image passed to the structural encoder. The stem is evaluated before the frozen encoder call, but the encoder itself is executed under \texttt{torch.no\_grad()}; its detached output therefore prevents the reconstruction objectives from propagating a gradient back to the stem. Although the stem parameters are declared trainable, the second convolution is zero-initialized and remains zero in the reported model state. Thus $g_{\mathrm{stem}}(\mathbf{X})=\mathbf{0}$ and $\mathbf{P}=\mathbf{I}_{\mathrm{obs}}$ for the reported model, so this nominal path is not a learned source of gain.

\paragraph{Token extraction.} Structural reasoning is instantiated with a frozen pretrained DINOv3 ViT-S/16 encoder~\cite{simeoni2025dinov3}. The encoder contains 12 transformer blocks, uses 384-dimensional tokens and six attention heads, and includes one class token and four register tokens. The padded pseudo-RGB input $\mathbf{P}$ is bicubically resized to $224\times224$, clipped to $[0,1]$, and normalized using the ImageNet mean and standard deviation. We retain the final normalized patch tokens and exclude the class and register tokens, yielding 196 patch tokens that are reshaped into $\mathbf{Z}\in\mathbb{R}^{384\times14\times14}$.

\paragraph{Feature adaptation and residual fusion.} The patch-token map is adapted and fused with the restoration representation as
\begin{align}
\mathbf{A}&=f_{\mathrm{adp}}(\operatorname{Resize}(\mathbf{Z})),\\
\boldsymbol{\Delta}&=f_{\mathrm{fuse}}([\mathbf{R},\mathbf{A}]),\\
\mathbf{R}_{f}&=\mathbf{R}+\boldsymbol{\Delta}.
\label{eq:supp_structural_fusion}
\end{align}
Here, $\operatorname{Resize}$ denotes bilinear interpolation to the internal $H_p\times W_p$ grid, $\mathbf{A}$ is the spatially and channel-aligned structural feature, $\boldsymbol{\Delta}$ is its restoration-dependent correction, and $\mathbf{R}_{f}$ is the fused representation. The adapter consists of a $1\times1$ convolution from 384 to 96 channels, GELU, and a $3\times3$ convolution. The structural fusion module consists of a $1\times1$ convolution from 192 to 96 channels, GELU, and a $3\times3$ convolution. The final fusion convolution is zero-initialized, so $\boldsymbol{\Delta}=\mathbf{0}$ and $\mathbf{R}_{f}=\mathbf{R}$ at initialization. This $192\!\rightarrow\!96$ residual fusion is distinct from the $288\!\rightarrow\!96$ sensor-stem fusion in Eq.~\eqref{eq:supp_initial_feature}; both output heads are cropped back to $H\times W$ after decoding.

\paragraph{Training and inference lifecycle.} The pretrained structural encoder remains in evaluation mode, all of its parameters are frozen, and its forward pass is executed without gradient storage. It nevertheless remains active in every training and inference pass. The restoration stream, adapter, residual fusion module, shared decoder, and output heads are optimized by the reconstruction objectives. The proposed online model uses neither clean-image structural targets nor a feature-matching or teacher--student distillation loss.

\subsection{Shared Decoder and Dual Output Heads}
\label{sec:supp_decoder}

The fused representation is first refined by a shared residual transform and then separated into RGB- and mosaic-specific residual transforms:
\begin{align}
\mathbf{D}_{s}&=\mathbf{R}_{f}+f_{\mathrm{shared}}(\mathbf{R}_{f}),\\
\mathbf{D}_{\mathrm{rgb}}&=\mathbf{D}_{s}+f_{\mathrm{rgb}}(\mathbf{D}_{s}),\\
\mathbf{D}_{\mathrm{mos}}&=\mathbf{D}_{s}+f_{\mathrm{mos}}(\mathbf{D}_{s}).
\label{eq:supp_decoder_features}
\end{align}
Here, $\mathbf{D}_{s}$ is the shared decoder feature, while $\mathbf{D}_{\mathrm{rgb}}$ and $\mathbf{D}_{\mathrm{mos}}$ are the RGB- and mosaic-specific features. All three transforms operate on 96 channels. Each of $f_{\mathrm{shared}}$, $f_{\mathrm{rgb}}$, and $f_{\mathrm{mos}}$ consists of a $3\times3$ convolution from 96 to 96 channels, GELU, and a second $3\times3$ convolution from 96 to 96 channels; the second convolution is zero-initialized, and no normalization is used. A $3\times3$ RGB head maps $\mathbf{D}_{\mathrm{rgb}}$ to three channels, while a separate $3\times3$ mosaic head maps $\mathbf{D}_{\mathrm{mos}}$ to one channel:
\begin{equation}
\widehat{\mathbf{I}}=H_{\mathrm{rgb}}(\mathbf{D}_{\mathrm{rgb}})+h_{\mathrm{skip}}(\mathbf{I}_{\mathrm{obs}}),\qquad \widehat{\mathbf{Y}}=H_{\mathrm{mos}}(\mathbf{D}_{\mathrm{mos}}).
\label{eq:supp_outputs}
\end{equation}
Here, $\widehat{\mathbf{I}}$ is the restored RGB image and $\widehat{\mathbf{Y}}$ is the auxiliary retained-mosaic prediction. The RGB skip mapping $h_{\mathrm{skip}}$ uses a $3\times3$ convolution from 3 to 48 channels, GELU, and a zero-initialized $3\times3$ convolution from 48 to 3 channels, preserving an observation-aligned residual path without imposing a fixed identity mapping. The auxiliary target is $\mathbf{Y}^{\mathrm{val}}$ in Eq.~\eqref{eq:supp_retained_mosaic}: it contains no sensor noise but follows the training-time retention mask and is zero at dropped locations. The mosaic branch regularizes the common fused representation during training, whereas only $\widehat{\mathbf{I}}$ is used for the reported RGB metrics.

\subsection{Training Objectives}
\label{sec:supp_objectives}

For an elementwise error $e$, the Charbonnier~\cite{charbonnier1994two} penalty is $\rho(e)=\sqrt{e^{2}+\epsilon^{2}}$ with $\epsilon=10^{-3}$, providing a smooth approximation to $|e|$. The three reconstruction objectives are
\begin{align}
\mathcal{L}_{\mathrm{rgb}}&=\operatorname{mean}_{c,u}\rho(\widehat{\mathbf{I}}_{c,u}-\mathbf{I}_{c,u}),\\
\mathcal{L}_{\mathrm{known}}&=\frac{\sum_{c,u}\mathbf{M}^{p,\mathrm{val}}_{c,u}\rho(\widehat{\mathbf{I}}_{c,u}-\mathbf{I}_{c,u})}{\sum_{c,u}\mathbf{M}^{p,\mathrm{val}}_{c,u}+10^{-8}},\\
\mathcal{L}_{\mathrm{mos}}&=\operatorname{mean}_{u}\rho(\widehat{\mathbf{Y}}_{u}-\mathbf{Y}^{\mathrm{val}}_{u}).
\label{eq:supp_losses}
\end{align}
The primary term $\mathcal{L}_{\mathrm{rgb}}$ averages reconstruction error over every RGB channel and spatial location. The term $\mathcal{L}_{\mathrm{known}}$ averages the same error only where a CFA measurement remains valid, with $10^{-8}$ preventing division by zero, while $\mathcal{L}_{\mathrm{mos}}$ supervises the auxiliary prediction against the clean retained mosaic. The complete objective is
\begin{equation}
\mathcal{L}_{\mathrm{total}}=\mathcal{L}_{\mathrm{rgb}}+0.25\,\mathcal{L}_{\mathrm{known}}+0.05\,\mathcal{L}_{\mathrm{mos}}.
\label{eq:supp_total_loss}
\end{equation}
The coefficients $0.25$ and $0.05$ are $\lambda_{\mathrm{known}}$ and $\lambda_{\mathrm{mos}}$, respectively; the implicit weight of $\mathcal{L}_{\mathrm{rgb}}$ is one. No perceptual, SSIM, structural feature-matching, distillation, or additional regularization loss is used for the proposed online model.

\subsection{End-to-End Data Flow}
\label{sec:supp_data_flow}

For clarity, one training iteration follows the same dependency order as the model equations:
\begin{enumerate}
\item Sample a clean crop, CFA family, noise level, and training-only measurement-dropout mask; use the fixed RGGB phase.
\item Form the five-channel sensor input $\mathbf{X}$ and sparse pseudo-RGB projection $\mathbf{I}_{\mathrm{obs}}$.
\item Produce the conditioned SwinIR restoration feature $\mathbf{R}$.
\item Extract frozen structural tokens from $\mathbf{P}$, adapt them to $\mathbf{A}$, and obtain the residual fusion $\mathbf{R}_{f}$.
\item Decode the shared representation into $\widehat{\mathbf{I}}$ and the auxiliary retained-mosaic prediction $\widehat{\mathbf{Y}}$.
\item Update only the non-frozen paths using Eq.~\eqref{eq:supp_total_loss}; no random measurement dropout is applied at standard inference.
\end{enumerate}

\section{Experimental Protocol}
\label{sec:supp_protocol}

\subsection{Datasets and Splits}
\label{sec:supp_datasets}

Training uses the 800 high-resolution training images from DIV2K~\cite{agustsson2017ntire}. The 18 McMaster images~\cite{zhang2011color} are reserved for validation and model selection. MIT-Hard is the union of the MIT-HDRVDP and MIT-Moir\'e test collections released with Deep Joint Demosaicking and Denoising~\cite{gharbi2016deep}; each collection contains 1,000 images. DIV2K provides diverse high-resolution content for online synthesis of multiple CFA and noise configurations, McMaster is a standard color-restoration benchmark, and MIT-Hard emphasizes the high-frequency and moir\'e-prone structures that are particularly diagnostic for demosaicing systems~\cite{gharbi2016deep,tedla2025examining}.

The original access pages are DIV2K\footnote{\url{https://data.vision.ee.ethz.ch/cvl/DIV2K/}}, McMaster\footnote{\url{https://www4.comp.polyu.edu.hk/~cslzhang/CDM_Dataset.htm}}, and the MIT dataset\footnote{\url{https://groups.csail.mit.edu/graphics/demosaicnet/dataset.html}}. Images are decoded as RGB, converted to \texttt{float32}, and normalized to $[0,1]$ without linearizing the decoded sRGB-like values. CFA sampling, noise, and measurement dropout are synthesized online from the clean images. Before evaluation synthesis, the bottom and right boundaries are cropped to the largest height and width divisible by the selected CFA period; no resizing is used. Thus a $500\times500$ image remains $500\times500$ for Single and Quad but becomes $498\times498$ for Nona. The supplementary package does not redistribute third-party images; the dataset names, citations, source URLs, expected image counts, and directory instructions are provided so that users can obtain them from the original sources.

The three dataset roles are kept disjoint: DIV2K supplies training content, McMaster controls model selection, and MIT-Hard is used only for final robustness analysis. The original access pages listed above therefore correspond to the training, validation, and MIT-Hard datasets used in the supplementary quantitative analyses.

\subsection{Training Configuration}
\label{sec:supp_training}

Training uses random $144\times144$ DIV2K crops, horizontal and vertical flips, and rotations by $0^\circ$, $90^\circ$, $180^\circ$, or $270^\circ$; Single-, Quad-, and Nona-Bayer families are sampled with equal probability. The crop size is divisible by the common CFA/window alignment of 24. Augmentation is applied to clean RGB content before a fixed-RGGB mask is regenerated, so these transforms do not sample additional CFA phases. The conditioned restoration stream is optimized before the extension stages, and the complete model is therefore not a 30-epoch from-scratch run: the dual-head stage runs for 50 epochs at learning rate $2\times10^{-5}$ with physical batch size two and four-step accumulation; the online structural-branch stage runs for 20 epochs at $10^{-5}$ with physical batch size one and eight-step accumulation; the shared-decoder stage runs for 20 epochs at $2\times10^{-5}$ with the same effective batch; and the final stage runs for 30 epochs with the cosine schedule~\cite{loshchilov2017sgdr} in Table~\ref{tab:supp_training_config}. The table therefore records the final fine-tuning stage, not the complete optimization history. Adam~\cite{kingma2015adam} uses $\beta_1=0.9$, $\beta_2=0.999$, $\epsilon=10^{-8}$, and zero weight decay. Training forward and backward passes do not use automatic mixed precision; gradient recomputation is enabled. Each final-stage epoch contains 800 micro-batches, corresponding to 100 optimizer updates after gradient accumulation. Balanced McMaster PSNR is used for model selection, and the same final model state is evaluated unchanged under every reported CFA family, noise level, and test subset.

\begin{table*}[!t]
\centering
\footnotesize
\setlength{\tabcolsep}{4.0pt}
\renewcommand{\arraystretch}{1.08}

\begin{tabular}{@{}lccclcccc@{}}
\toprule
Crop
& Batch
& Epochs
& Optimizer
& LR schedule
& Noise
& Dropout
& $\lambda_{\mathrm{known}}$
& $\lambda_{\mathrm{mos}}$ \\
\midrule
$144\times144$
& $1\times8$
& 30
& Adam
& $10^{-5}\!\rightarrow\!10^{-6}$ cosine
& $\mathcal{U}(0,0.098)$
& $\mathcal{U}(0,0.05)$
& 0.25
& 0.05 \\
\bottomrule
\end{tabular}

\caption{Configuration of the final 30-epoch joint fine-tuning stage of the online model. ``$1\times8$'' denotes a physical batch size of one with eight-step gradient accumulation. The random seed is 1234 and weight decay is zero.}
\label{tab:supp_training_config}
\end{table*}

\subsection{Model Selection and Evaluation Protocol}
\label{sec:supp_evaluation}

Balanced McMaster validation PSNR determines model selection during the final optimization stage; no MIT-Hard image is used for this decision. Final benchmark tables use standalone FP32 inference without tiling, a two-pixel border crop, seed 1234, RGGB phase, no test-time measurement dropout, and the 8-bit metric conversion described below. A single validation-selected model state is used for every reported online-model MIT-Hard noise level, CFA family, and test subset, with no condition-specific tuning.

Inference is performed in FP32, but the metric inputs are quantized to 8-bit RGB: predictions and targets are clipped to $[0,1]$, multiplied by 255, rounded, and converted to \texttt{uint8}. After the explicit two-pixel border crop, PSNR is computed jointly over all retained spatial locations and three color channels as $20\log_{10}(255/\sqrt{\mathrm{MSE}_{8\mathrm{bit}}})$. SSIM~\cite{wang2004ssim} is computed independently on each RGB channel and then averaged, using an $11\times11$ Gaussian window with standard deviation 1.5, $C_1=(0.01\cdot255)^2$, $C_2=(0.03\cdot255)^2$, and valid-window averaging. Each metric is computed per image before arithmetic averaging. Accordingly, ``FP32 evaluation'' below means FP32 inference with 8-bit-quantized metric inputs, not direct metric computation on floating-point predictions.

McMaster validation enumerates Single-, Quad-, and Nona-Bayer observations at four noise levels: $0$, $5/255$, $15/255$, and $25/255$. Their implemented decimal values are $0$, $0.0196$, $0.0588$, and $0.098$, respectively. If $m_{i,p,\sigma}$ denotes a per-image PSNR or SSIM value, the balanced metric is
\begin{equation}
m_{\mathrm{bal}}=\frac{1}{|\mathcal{P}||\Sigma|}\sum_{p\in\mathcal{P}}\sum_{\sigma\in\Sigma}\frac{1}{N_{\mathrm{img}}}\sum_{i=1}^{N_{\mathrm{img}}}m_{i,p,\sigma}.
\label{eq:supp_balanced}
\end{equation}
Here, $\mathcal{P}$ is the three-family CFA set defined above, $\Sigma=\{0,5/255,15/255,25/255\}$, and $N_{\mathrm{img}}=18$. Averaging first over images and then equally over the $|\mathcal{P}||\Sigma|=12$ CFA--noise groups gives each group the same weight.

MIT-Hard reports nonzero-noise performance at $\sigma\in\{5/255,15/255,25/255\}$ for both 1,000-image subsets and all three CFA families. Each subset therefore contains 9,000 evaluations per method, and the combined nonzero-noise result contains 18,000 evaluations per method. A single fixed noise realization is used for each image--CFA--noise case rather than averaging over repeated noise draws. All compared methods use the same synthesized observations, fixed RGGB phase, noise realizations, border crop, quantization, and metric implementation.

The reporting hierarchy separates three kinds of evidence. Aggregate MIT-Hard values summarize overall robustness, the complete subset--CFA--noise matrix reveals condition-specific behavior, and paired McMaster differences test whether an average validation gain is broadly distributed across cases. Model-development results then provide within-family evidence for the progressive architecture, whereas the cross-method table provides an external comparison under a shared evaluator. Because McMaster is also used for model selection, its paired analysis is a validation-set diagnostic rather than independent held-out test evidence. These roles are complementary and should not be treated as interchangeable forms of ablation.

\subsection{Compared Methods and Scope}
\label{sec:supp_comparisons}

The main paper compares the proposed model with LSUM, SRUM, ESUM, and a unified All-in-One SwinIR baseline. These methods retain their architecture-specific representations and training recipes, while the observation generator and evaluator are shared. The comparison is therefore controlled at the data and evaluation levels rather than being a matched-parameter, matched-loss, or matched-training-budget study. In particular, the All-in-One entry in Table~\ref{tab:supp_mit_hard} is the original four-channel noise-blind configuration, which receives the mosaic and three CFA masks, omits the noise map, and was trained only on noise-free synthesized observations; ESUM was likewise trained at zero noise. Their nonzero-noise results therefore measure out-of-distribution robustness without retraining and must not be interpreted as a matched test of the online structural branch.

\section{Expanded Results and Analysis}
\label{sec:supp_results}
\suppressfloats[t]

The following analyses move from dataset-level averages to condition-level and case-level evidence. This ordering is intentional: a single average is concise but can obscure behavior on the harder CFA families or at the largest noise level, while isolated visual crops cannot establish dataset-wide consistency. Reporting both views gives a more complete account of what the existing experiments do and do not support.

\subsection{Supplementary MIT-Hard Diagnostic}
\label{sec:supp_mit_hard}

Tables~\ref{tab:supp_mit_hard} and~\ref{tab:supp_mit_matrix} report an additional standalone evaluation of the online architecture under the protocol in Sec.~\ref{sec:supp_evaluation}. The first table gives the nonzero-noise average over the two MIT-Hard subsets, three CFA families, and three nonzero noise levels; the second exposes every subset--CFA--noise group. This diagnostic is included to support condition-wise analysis and is not used to replace or recompute the aggregate reported in the main paper.

\begin{table}[!h]
\centering
\footnotesize
\setlength{\tabcolsep}{5.0pt}
\renewcommand{\arraystretch}{1.08}
\begin{tabular}{@{}lcc@{}}
\toprule
Method & PSNR (dB) & SSIM \\
\midrule
All-in-One SwinIR & 23.9559 & 0.6741 \\
ESUM / RCAN & 24.2447 & 0.6770 \\
\textbf{Ours} & \textbf{27.9629} & \textbf{0.8368} \\
\bottomrule
\end{tabular}
\caption{Supplementary standalone MIT-Hard diagnostic averaged over MIT-HDRVDP and MIT-Moir\'e, Single-, Quad-, and Nona-Bayer observations, and $\sigma\in\{5/255,15/255,25/255\}$.}
\label{tab:supp_mit_hard}
\end{table}

\begin{table}[!b]
\centering
\scriptsize
\setlength{\tabcolsep}{2.5pt}
\renewcommand{\arraystretch}{1.08}
\begin{tabular}{@{}llcccc@{}}
\toprule
Subset & CFA & $\sigma=0$ & $\sigma=5/255$ & $\sigma=15/255$ & $\sigma=25/255$ \\
\midrule
\multirow{3}{*}{MIT-HDRVDP} & Single & 32.3741/.9489 & 30.8809/.9288 & 27.3954/.8591 & 25.2235/.7850 \\
& Quad & 30.2896/.9373 & 29.3192/.9180 & 26.6333/.8491 & 24.7883/.7765 \\
& Nona & 29.6437/.9320 & 28.7220/.9120 & 26.1812/.8420 & 24.4327/.7688 \\
\midrule
\multirow{3}{*}{MIT-Moir\'e} & Single & 34.3598/.9289 & 32.3492/.9044 & 28.8728/.8278 & 26.9116/.7583 \\
& Quad & 33.1499/.9235 & 31.4138/.8992 & 28.3547/.8208 & 26.5794/.7507 \\
& Nona & 32.3732/.9202 & 30.8862/.8956 & 28.0460/.8174 & 26.3424/.7480 \\
\bottomrule
\end{tabular}
\caption{Group matrix for the supplementary online-model diagnostic, reported as RGB PSNR (dB) / SSIM. Each entry averages 1,000 images under the same FP32, RGGB, crop-border-2 evaluation protocol. The $\sigma=0$ column is shown for completeness but is excluded from Table~\ref{tab:supp_mit_hard}.}
\label{tab:supp_mit_matrix}
\end{table}

The expanded matrix shows three consistent trends. First, PSNR and SSIM decrease monotonically with increasing noise in all six subset--CFA groups, so the aggregate is not produced by averaging conflicting trends. Second, Single-Bayer is consistently the least difficult family and Nona-Bayer the most difficult within each subset and noise level, consistent with larger same-color blocks and longer distances to complementary color samples. Third, both listed noise-free-trained baselines are substantially less robust under this nonzero-noise diagnostic. As stated in Sec.~\ref{sec:supp_comparisons}, these comparisons characterize robustness under a common evaluator rather than the isolated causal contribution of structural guidance.


\subsection{Model-Development Analysis}
\label{sec:supp_model_development}

\begin{table}[!h]
\centering
\footnotesize
\setlength{\tabcolsep}{1.8pt}
\renewcommand{\arraystretch}{1.08}
\begin{tabular}{@{}lccc@{}}
\toprule
Variant & Bal. & $\sigma=0$ & $\sigma=25/255$ \\
\midrule
Dual-head & 32.461 & 35.637 & 29.087 \\
$+$ Frozen structural encoder & 32.477 & 35.651 & 29.101 \\
$+$ Shared trunk & 32.505 & 35.696 & 29.114 \\
$+$ Fine-tuning & 32.525 & 35.726 & 29.137 \\
\midrule
\textbf{Full model} & \textbf{32.732} & \textbf{36.096} & \textbf{29.249} \\
\bottomrule
\end{tabular}
\caption{Model-development analysis on McMaster in PSNR (dB). Balanced denotes the average over Single-, Quad-, and Nona-Bayer observations at all four evaluated noise levels. The Full model is listed as the separately reported final configuration.}
\label{tab:supp_model_analysis}
\end{table}

The first four rows of Table~\ref{tab:supp_model_analysis} summarize the development from the dual-head restoration model to the frozen structural branch, shared decoding, and subsequent fine-tuning. Across these four configurations, balanced PSNR increases monotonically by 0.064~dB in total; the cumulative gains at $\sigma=0$ and $\sigma=25/255$ are 0.089~dB and 0.050~dB, respectively. Thus the development trend is present in both the clean and strongest-noise columns rather than being confined to one operating point, and the separately reported Full model is highest in all three columns. The rows are sequential complete configurations rather than a factorial leave-one-component-out study, so the increments support consistency of the development path but should not be interpreted as isolated causal effects for individual components.

\subsection{Paired Validation Diagnostic on McMaster}
\label{sec:supp_paired_stability}

Aggregate means can be dominated by a small number of easy or difficult images, so we additionally compare methods case by case over all $18\times3\times4=216$ McMaster image--CFA--noise combinations. Under standalone FP32 inference with the crop-border-2 and 8-bit metric protocol, the online Full model exceeds both listed noise-free-trained baselines in every paired case. Table~\ref{tab:supp_paired_stability} reports the distribution of PSNR differences rather than only their overall mean. Because McMaster is used for model selection, these statistics are a validation-set diagnostic, not independent test evidence; they also inherit the comparison scope in Sec.~\ref{sec:supp_comparisons} and are not a matched structural-guidance ablation.

\begin{table*}[!t]
\centering
\footnotesize
\setlength{\tabcolsep}{5.0pt}
\renewcommand{\arraystretch}{1.08}

\begin{tabular}{@{}lccccc@{}}
\toprule
Comparison & Cases & Win rate & Mean gain & Median gain & Gain range \\
\midrule
Full model $-$ All-in-One
& 216
& 100\%
& 5.6710 dB
& 5.1775 dB
& $[1.7802,\,13.0035]$ dB \\

Full model $-$ ESUM
& 216
& 100\%
& 4.8699 dB
& 4.3639 dB
& $[0.7697,\,13.0708]$ dB \\
\bottomrule
\end{tabular}

\caption{Paired validation-set diagnostic on McMaster over 18 images, three CFA families, and four noise levels. Positive values favor the online Full model.}
\label{tab:supp_paired_stability}
\end{table*}

The positive lower endpoints of both gain ranges are important: even the least improved paired cases remain above the corresponding listed baseline. The median gains are smaller than the means, indicating that several large improvements raise the average, but the 100\% win rates show that those large cases are not the sole source of the aggregate advantage. This case-level view therefore complements, rather than duplicates, the dataset-level averages.

\subsection{Residual Adaptation and Shared Decoding}
\label{sec:supp_design_analysis}

The structural fusion is residual by construction. The 384-dimensional token map is adapted to the 96-dimensional restoration space, concatenated with $\mathbf{R}$, and converted into the correction $\boldsymbol{\Delta}$ in Eq.~\eqref{eq:supp_structural_fusion}. Zero initialization of the last fusion convolution makes the initial mapping exactly $\mathbf{R}_{f}=\mathbf{R}$, after which optimization can learn where structural context should modify the sensor-aware representation. The shared decoder then exposes a common refined feature to both reconstruction objectives before the RGB and mosaic paths separate, allowing the auxiliary sensor-domain target to regularize shared content without forcing every RGB-specific decoder feature to serve the mosaic task. This design interpretation is consistent with the monotonic development trend in Table~\ref{tab:supp_model_analysis}, while the sequential nature of that table prevents assigning the full gain to any one module in isolation.

\subsection{Implementation Distinction from DINOv2 Distillation}
\label{sec:supp_dinov2}

To distinguish training-only feature supervision from the proposed online structural branch, we also consider a DINOv2 feature-distillation baseline~\cite{oquab2024dinov2}. During that experiment, a frozen DINOv2 ViT-S/14 teacher bicubically resizes the clean RGB target to $224\times224$ and produces a $384\times16\times16$ feature map. A student head applies a $1\times1$ convolution from 96 to 384 channels to the restoration representation and adaptive average pooling to $16\times16$. With $N_{\mathrm{tok}}=16\times16$ spatial tokens and feature dimension $C_{\mathrm{feat}}=384$, the default mean-squared-error reduction implemented in code corresponds to
\begin{equation}
\mathcal{L}_{\mathrm{dist}}=\frac{1}{N_{\mathrm{tok}}C_{\mathrm{feat}}}\sum_{t=1}^{N_{\mathrm{tok}}}\left\|\frac{\widehat{\mathbf{v}}_{t}}{\|\widehat{\mathbf{v}}_{t}\|_{2}}-\frac{\mathbf{v}_{t}}{\|\mathbf{v}_{t}\|_{2}}\right\|_{2}^{2},
\label{eq:supp_dinov2_loss}
\end{equation}
where $\mathbf{v}_{t}$ and $\widehat{\mathbf{v}}_{t}$ are the teacher feature vector and its student prediction at token $t$, respectively, and $\|\cdot\|_{2}$ is the Euclidean norm. Unit normalization removes feature magnitude, and division by $C_{\mathrm{feat}}$ reflects that \texttt{mse\_loss} averages over both spatial tokens and feature channels; the loss is added only in this experiment with weight 0.01. The teacher is used only during training, and the student head, although it can remain stored in the model object, is not executed in the ordinary RGB inference path. This comparison was trained from scratch for 400 epochs without the dual-head or shared-decoder components, so it is an implementation contrast rather than a matched ablation. By contrast, the proposed online branch processes the noisy sparse observation, remains active at inference, and uses no feature-distillation objective; the DINOv2 experiment is therefore not the Full model. Table~\ref{tab:supp_dino_comparison} is included to prevent these two uses of pretrained representations from being conflated, not to claim a matched performance ranking between them.

\begin{table}[!h]
\centering
\footnotesize
\setlength{\tabcolsep}{3.5pt}
\renewcommand{\arraystretch}{1.08}
\begin{tabular}{@{}lll@{}}
\toprule
Property & DINOv2 distillation & Online structural branch \\
\midrule
Teacher input & clean RGB & none \\
Structural input & training target & noisy sparse observation \\
Feature loss & yes & no \\
Encoder at inference & removed & active and frozen \\
Dual-domain decoder & absent in this baseline & present \\
\bottomrule
\end{tabular}
\caption{Conceptual distinction between the training-only DINOv2 feature-distillation comparison and the proposed online structural reasoning branch.}
\label{tab:supp_dino_comparison}
\end{table}

\subsection{Parameters, Computation, and Runtime}
\label{sec:supp_complexity}

Table~\ref{tab:supp_parameter_breakdown} separates the frozen structural encoder from the optimized restoration components. The nominal trainable count includes 3,507 pseudo-RGB stem parameters. As explained in Sec.~\ref{sec:supp_structural}, the stem is evaluated before the no-gradient encoder call, but the detached encoder output prevents the reconstruction objectives from reaching those parameters. Subtracting this nominal path gives 3,392,455 parameters that can receive reconstruction gradients in the reported configuration.

\begin{table}[!h]
\centering
\scriptsize
\setlength{\tabcolsep}{3.0pt}
\renewcommand{\arraystretch}{1.08}
\resizebox{\columnwidth}{!}{%
\begin{tabular}{@{}lrr@{}}
\toprule
Module & Total parameters & \texttt{requires\_grad=True} \\
\midrule
Conditioned stems and modulation & 426,387 & 426,387 \\
SwinIR body & 2,242,800 & 2,242,800 \\
DINOv3 encoder & 21,601,152 & 0 \\
Adapter & 120,000 & 120,000 \\
Pseudo-RGB stem and residual fusion & 105,075 & 105,075 \\
Shared decoder trunk & 166,080 & 166,080 \\
RGB decoder and head & 168,675 & 168,675 \\
Mosaic decoder and head & 166,945 & 166,945 \\
\midrule
\textbf{Full model} & \textbf{24,997,114} & \textbf{3,395,962} \\
\bottomrule
\end{tabular}%
}
\caption{Parameter breakdown of the online Full model. The right column is a parameter-flag count rather than a guarantee that every listed parameter receives a gradient in the reported configuration.}
\label{tab:supp_parameter_breakdown}
\end{table}

\begin{table}[!h]
\centering
\footnotesize
\setlength{\tabcolsep}{4.0pt}
\renewcommand{\arraystretch}{1.08}
\begin{tabular}{@{}ll@{}}
\toprule
Setting & Value \\
\midrule
Total parameters & 24.997M \\
Nominal trainable parameters & 3.396M \\
Core computation & 1912.08 GFLOPs \\
FP32 runtime & $1287.95\pm49.08$ ms \\
Peak allocated memory & 2077.86 MiB \\
\bottomrule
\end{tabular}
\caption{Complexity measured for a $1\times5\times512\times512$ input in FP32 without tiling. Runtime and peak allocated memory were measured on an RTX~3090 after 20 warm-up iterations and over 100 synchronized CUDA-event iterations. GFLOPs use $1$ MAC $=2$ FLOPs and count the core convolution, linear, and attention operations in the conditioned stem, SwinIR body, DINOv3 encoder, adapter, fusion, shared decoder, RGB branch, and mosaic branch; normalization, softmax, activation, interpolation, and elementwise operations are not included.}
\label{tab:supp_complexity}
\end{table}

The reported $\pm49.08$~ms is the population standard deviation over the 100 synchronized timed iterations. Runtime includes the frozen online structural encoder and both output branches; the FLOP total follows the operation exclusions stated in the caption. The frozen encoder accounts for approximately 86.4\% of all parameters, leaving 13.6\% nominally trainable. Freezing therefore keeps the number of updated parameters small, but it does not eliminate inference computation because structural tokens are extracted online. The complexity values should accordingly be read as the end-to-end cost of the proposed accuracy--structure trade-off, not as the cost of the restoration backbone alone.

\section{Reproducibility Guide}
\label{sec:supp_reproducibility}

\subsection{Software Environment and Randomness}
\label{sec:supp_environment}

The reported environment uses Python 3.10.20, PyTorch 2.5.1 with CUDA 12.1 support, torchvision 0.20.1, and cuDNN 9.1.0. The training program sets seed 1234 for Python, NumPy, PyTorch CPU, and all CUDA devices. The DataLoader does not define an explicit \texttt{generator} or \texttt{worker\_init\_fn}; its worker seeds therefore follow PyTorch's implicit base-seed mechanism rather than a separately documented DataLoader seed. Deterministic algorithms and deterministic cuDNN execution are not forced, and random-number-generator states are not explicitly restored after interrupted training; the seed controls the intended stochastic sources, but bitwise-identical training is not guaranteed across different GPU, CUDA, cuDNN, or PyTorch environments.

\subsection{Evaluation Reproduction}
\label{sec:supp_evaluation_reproduction}

Exact reproduction requires the conditioned SwinIR stream, DINOv3 adapter and residual fusion, shared decoder trunk, RGB decoder and head, and mosaic decoder and head, with no distillation head in the online model. The frozen DINOv3 encoder is acquired separately from its original source and loaded before the remaining trained model state. The accompanying README identifies the structural encoder source, pretrained weights, architecture flags, and evaluation configuration so that the intended online path can be instantiated without ambiguity.

The reproducible evaluation order is: acquire the public datasets and pretrained structural weights from their original sources; crop each image to the selected CFA period; construct the five-channel observation with the fixed RGGB phase, stated noise seed, and noise-map scaling; instantiate the online model with both the structural branch and shared decoder enabled; load the frozen encoder separately and strictly validate all non-encoder keys in the trained model state; run FP32 inference without tiling unless otherwise stated; clip, round, and convert predictions and targets to 8-bit RGB; crop two boundary pixels; compute per-image RGB PSNR and channel-averaged SSIM; and finally average the per-image metrics within each reported group. This order keeps data synthesis, inference, metric conversion, and aggregation explicit, which is necessary because changing any one of them can shift the reported values.

\subsection{Supplementary Code and Data Boundary}
\label{sec:supp_code_release}

The accompanying code archive provides a minimal executable demonstration rather than a redistribution of the full training repository. It contains an anonymous README, the online model definition, observation-synthesis code, a representative inference path, the learned non-encoder model state, strict validation of all non-encoder keys, the standalone evaluator, the configuration values reported here, and directory instructions. In particular, an implementation that accepts but silently ignores \texttt{enable\_dinov3\_branch} or \texttt{enable\_shared\_decoder\_trunk} is incompatible with the reported architecture. Third-party datasets and pretrained structural weights are obtained from their original sources; the archive supplies the scripts and instructions needed to reconstruct the evaluation inputs. The reproducibility claims made during review are limited to this document and the files included in the anonymous submission.

\section{Limitations, Scope, and Ethics}
\label{sec:supp_limitations}

\subsection{Limitations and Scope}

The current study synthesizes mosaics from decoded sRGB images and assumes independent signal-independent Gaussian noise. It therefore does not model heteroscedastic RAW noise, sensor-specific calibration, or a complete camera image-signal-processing pipeline. Training and evaluation focus on fixed-RGGB Single-, Quad-, and Nona-Bayer observations; generalization to other phases and CFA layouts remains to be measured. Each evaluation case uses one fixed noise realization, and test-time measurement-dropout robustness has not been measured. The frozen structural encoder adds online inference cost and total parameters, its fixed $224\times224$ input stretches non-square observations before token extraction, and DINOv3 ViT-S/16 is the only online structural encoder evaluated. The model-development results come from single runs and do not provide multi-seed uncertainty estimates, while the cross-method nonzero-noise comparison is not a matched training study. The learned reconstruction may also smooth or hallucinate fine detail when the CFA observation is severely undersampled or when repetitive texture is indistinguishable from aliasing and noise.

\subsection{Ethical Considerations}

The work uses public image-restoration benchmarks and involves no human participants or personal data collection. As with other learned reconstruction systems, the output may contain plausible but inaccurate fine detail in severely undersampled or noisy regions; restored images should therefore not be treated as unqualified evidence in forensic, medical, or other high-stakes settings.


\section{Additional Visual Comparisons}
\label{sec:supp_visual_appendix}

\begin{figure*}[!t]
\centering
\setlength{\tabcolsep}{0.35pt}
\renewcommand{\arraystretch}{1.0}

\newcommand{\MethodPanel}[1]{%
    \includegraphics[width=\linewidth]{#1}%
}

\newcommand{\CFARowLabel}[1]{%
    \rotatebox[origin=c]{90}{%
        \strut\scriptsize\bfseries #1%
    }%
}

\begin{tabular}{
    @{}
    >{\centering\arraybackslash}m{0.025\textwidth}
    @{\hspace{0.4mm}}
    *{5}{>{\centering\arraybackslash}m{0.190\textwidth}}
    @{}
}

&
\scriptsize LSUM
&
\scriptsize ESUM
&
\scriptsize All-in-One
&
\scriptsize Ours
&
\scriptsize Ground Truth
\\[-0.2mm]

\CFARowLabel{Single}
&
\MethodPanel{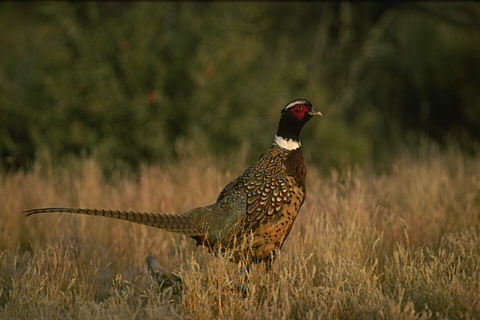}
&
\MethodPanel{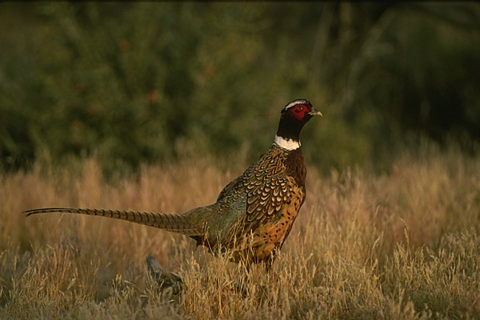}
&
\MethodPanel{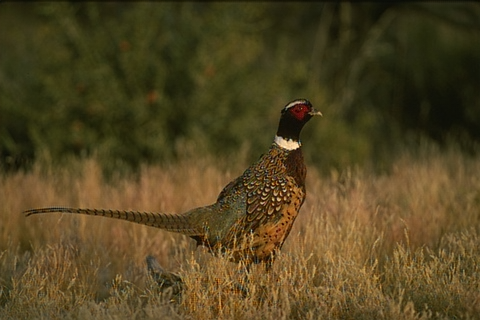}
&
\MethodPanel{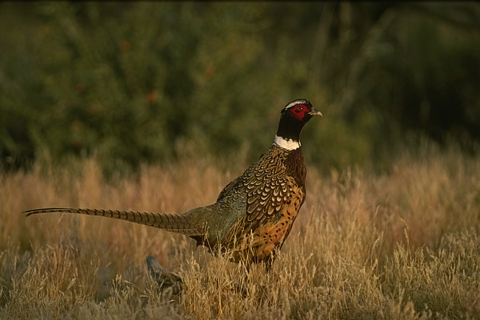}
&
\MethodPanel{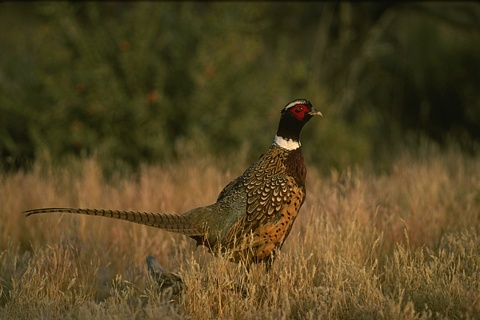}
\\[-0.7mm]

\CFARowLabel{Quad}
&
\MethodPanel{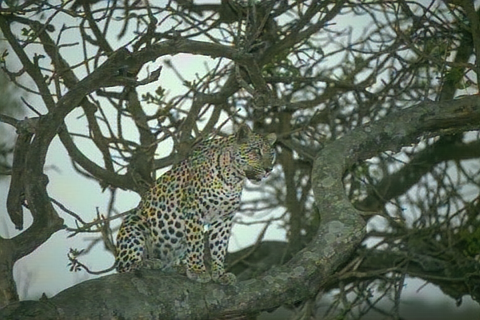}
&
\MethodPanel{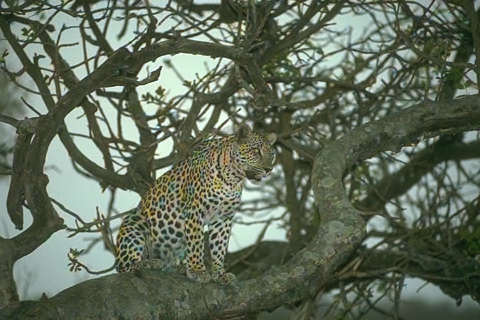}
&
\MethodPanel{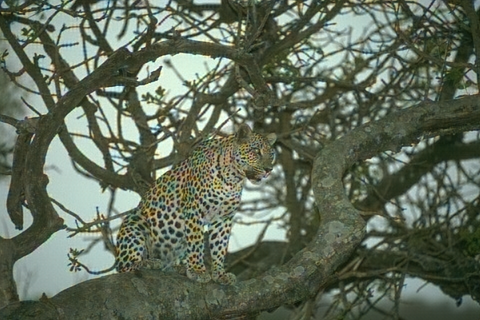}
&
\MethodPanel{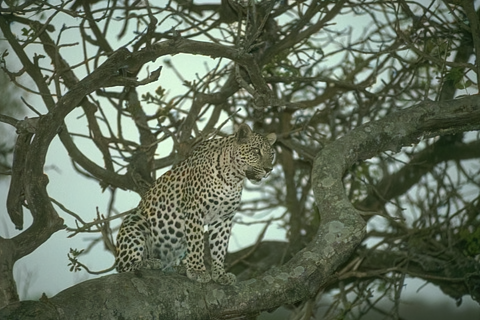}
&
\MethodPanel{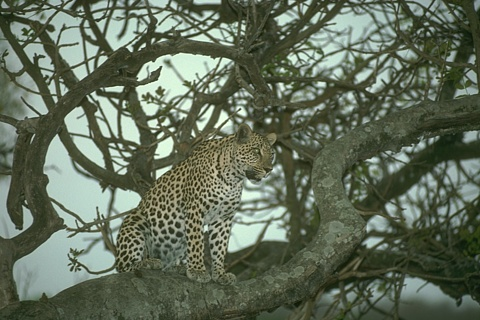}
\\[-0.7mm]

\CFARowLabel{Nona}
&
\MethodPanel{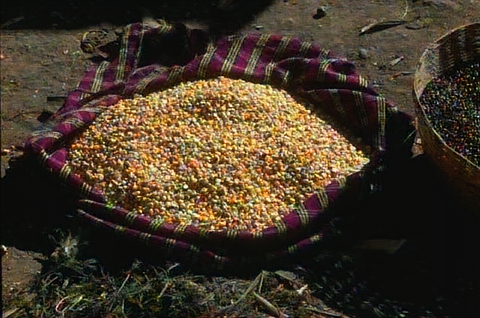}
&
\MethodPanel{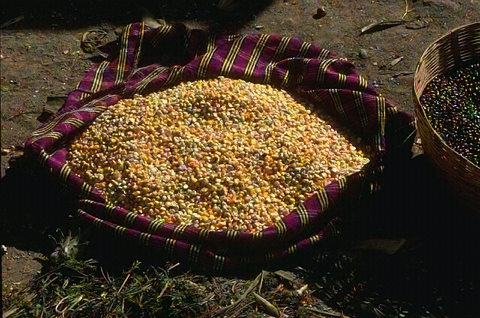}
&
\MethodPanel{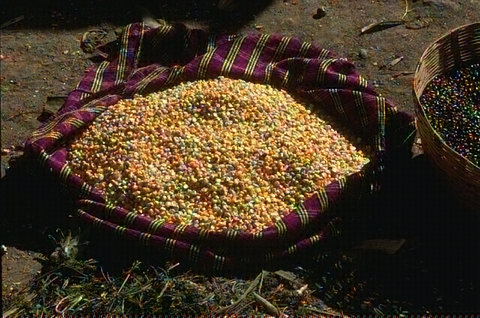}
&
\MethodPanel{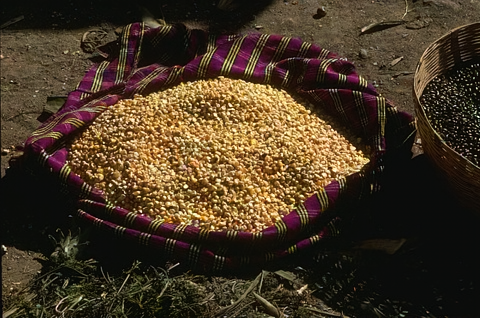}
&
\MethodPanel{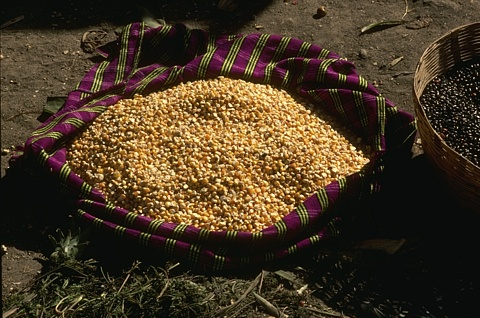}
\\[-0.3mm]

&
\multicolumn{5}{c}{
    \scriptsize\textit{BSD100}~\cite{martin2001database}
}
\\[0.5mm]

\CFARowLabel{Single}
&
\MethodPanel{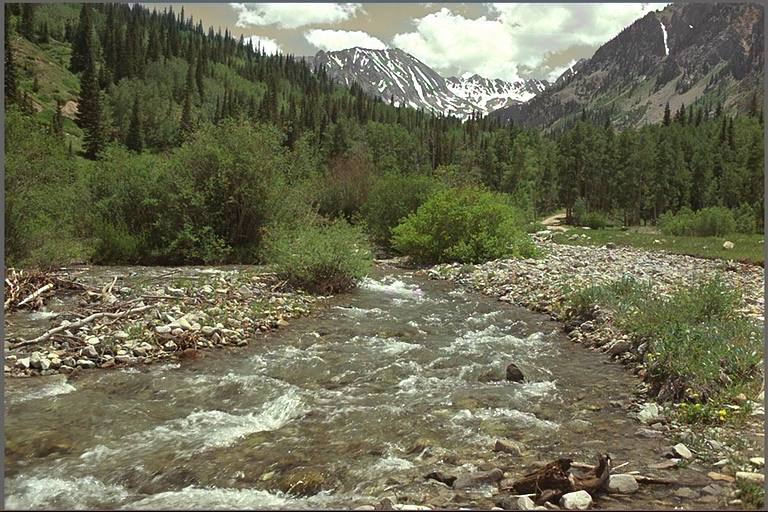}
&
\MethodPanel{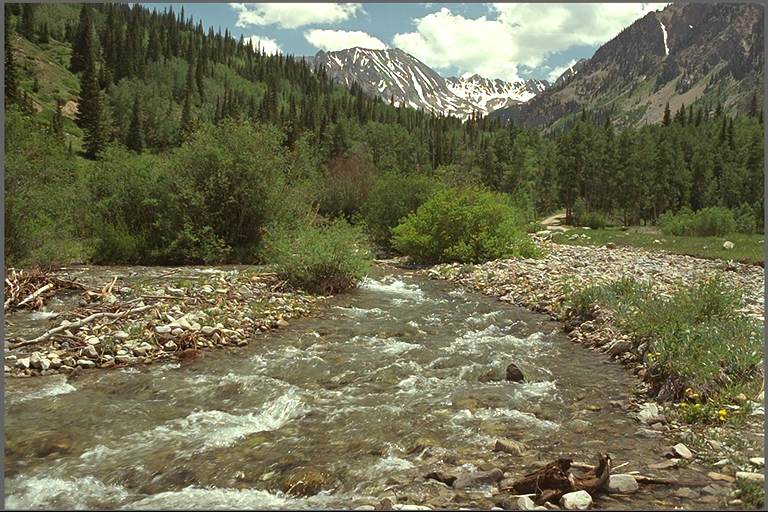}
&
\MethodPanel{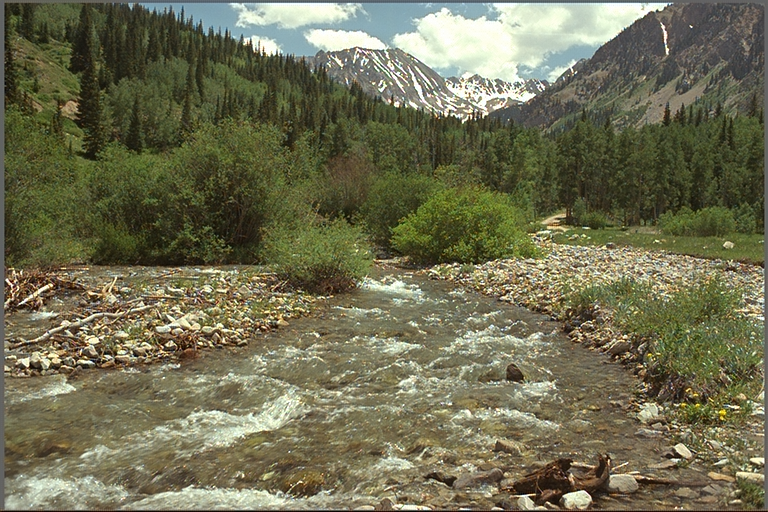}
&
\MethodPanel{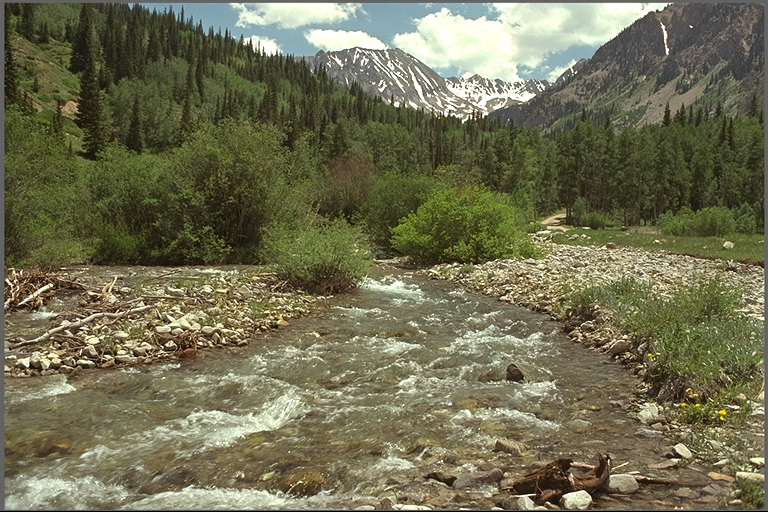}
&
\MethodPanel{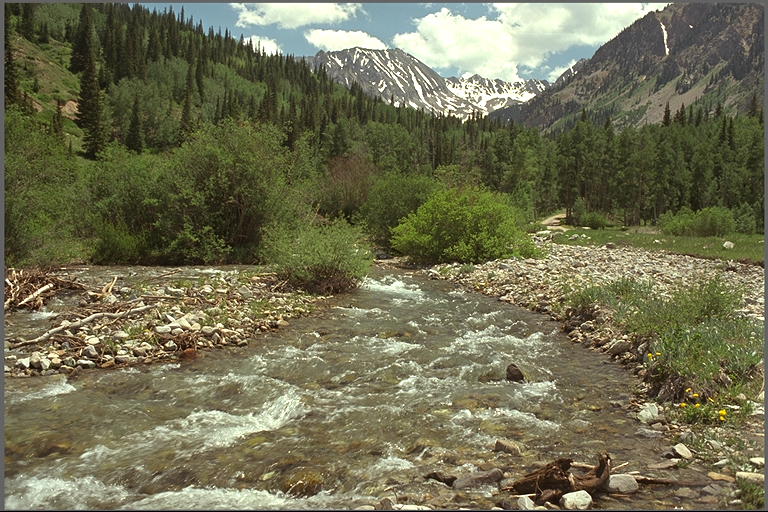}
\\[-0.7mm]

\CFARowLabel{Quad}
&
\MethodPanel{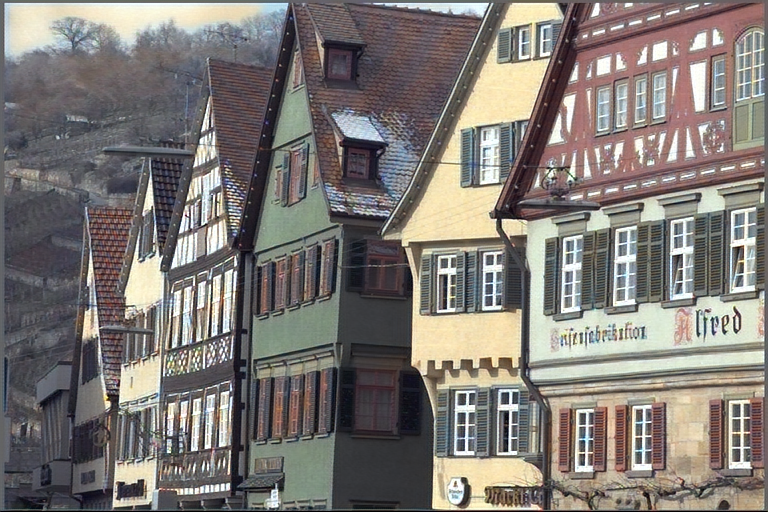}
&
\MethodPanel{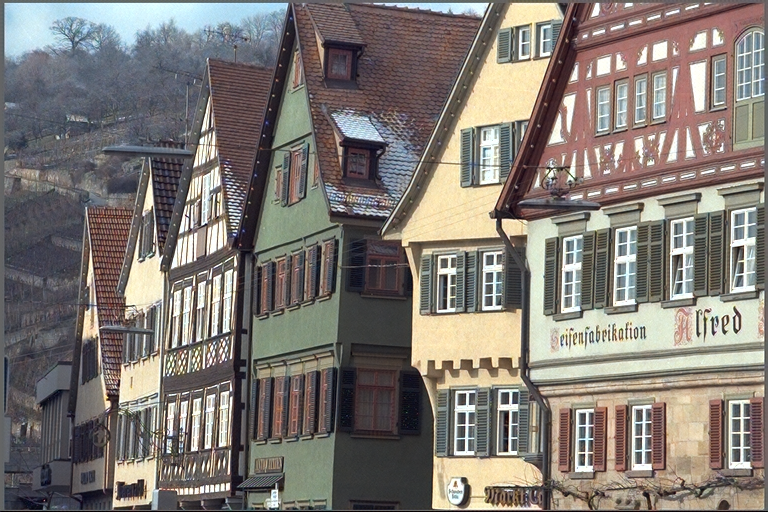}
&
\MethodPanel{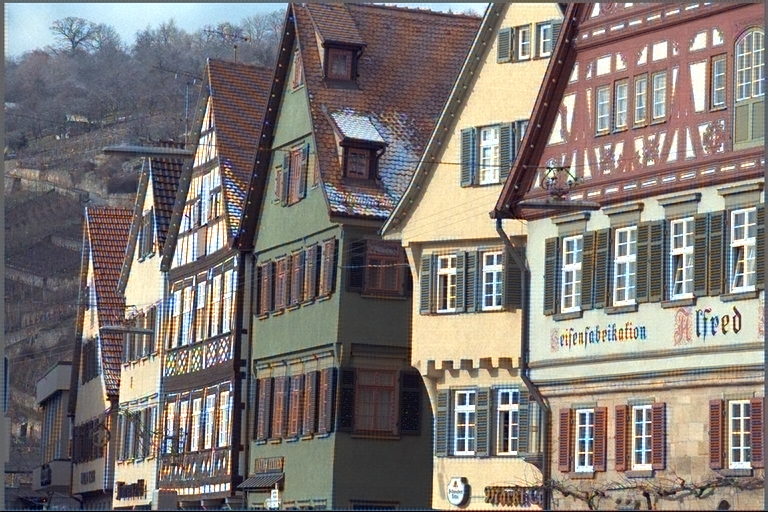}
&
\MethodPanel{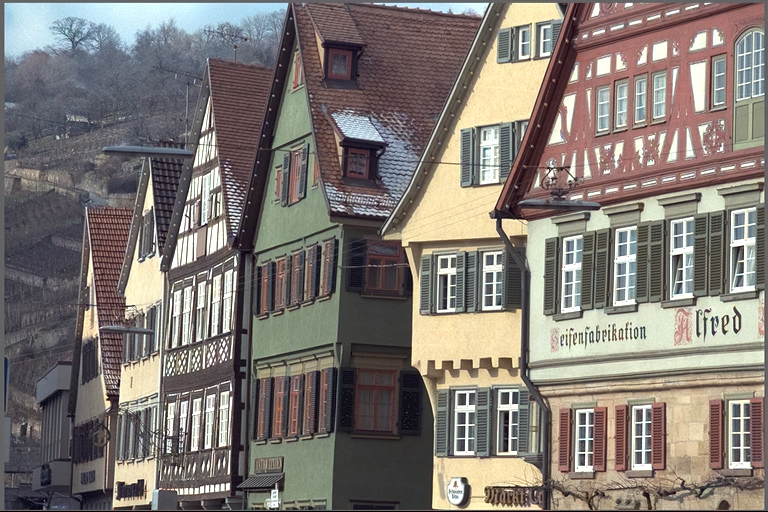}
&
\MethodPanel{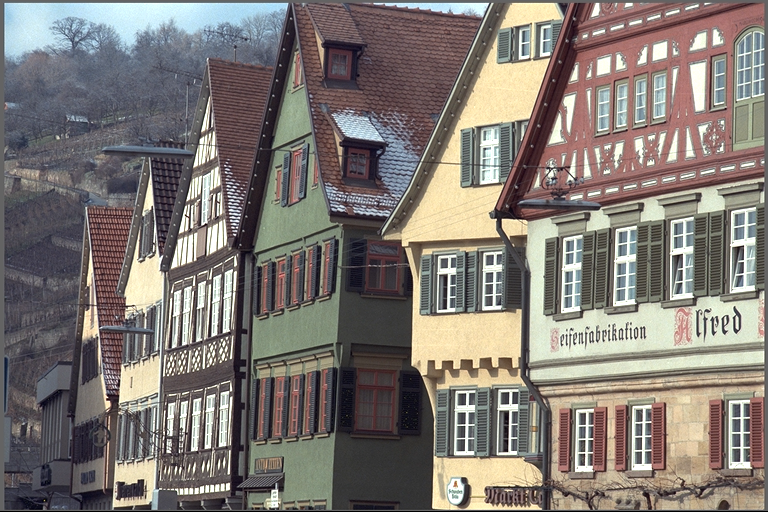}
\\[-0.7mm]

\CFARowLabel{Nona}
&
\MethodPanel{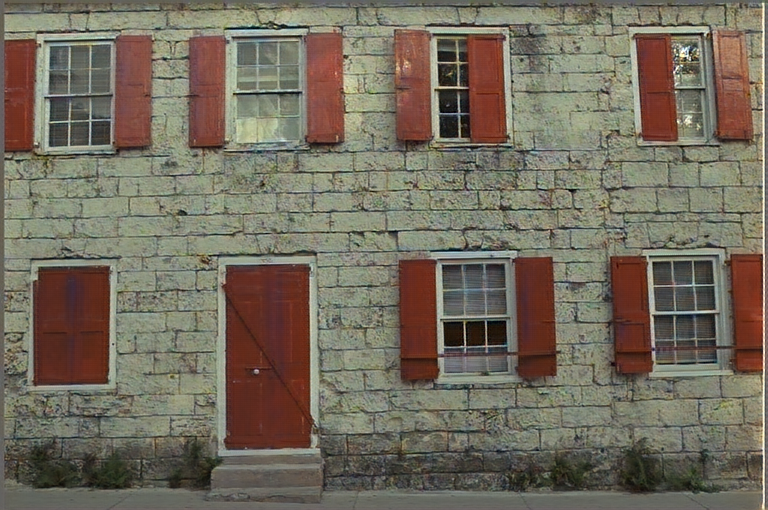}
&
\MethodPanel{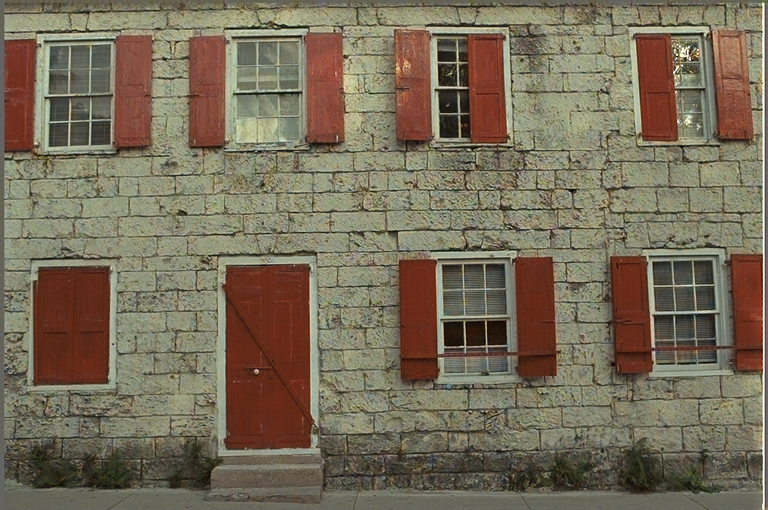}
&
\MethodPanel{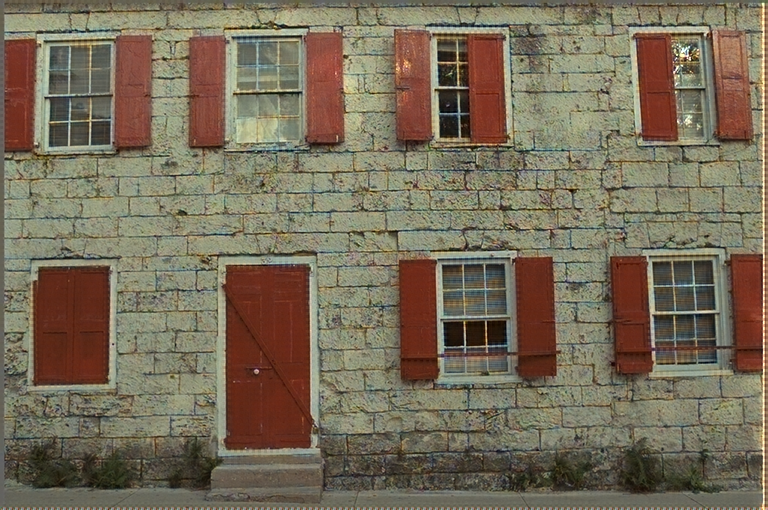}
&
\MethodPanel{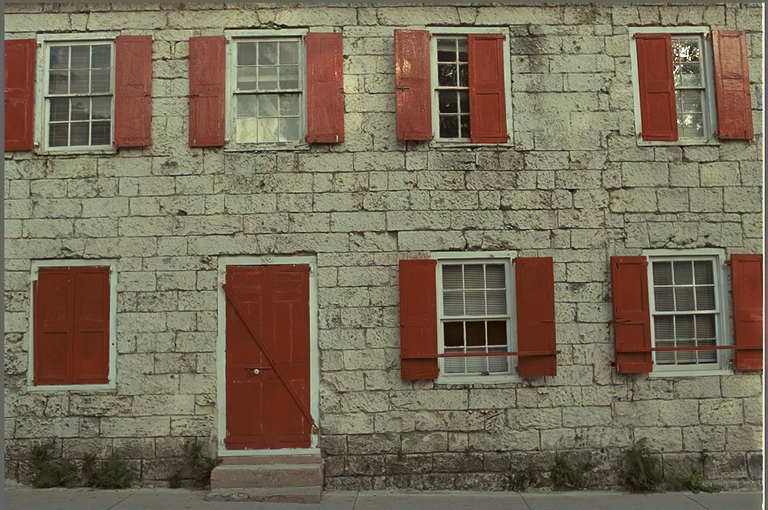}
&
\MethodPanel{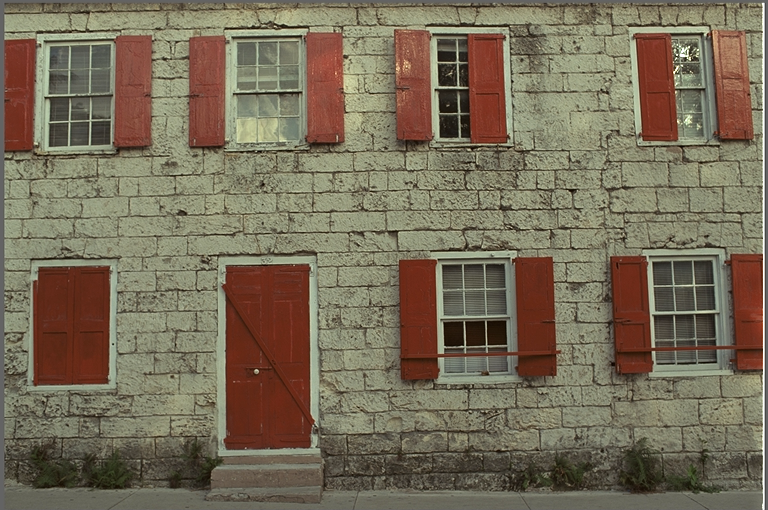}
\\[-0.3mm]

&
\multicolumn{5}{c}{
    \scriptsize\textit{Kodak24}~\cite{kodak24}
}
\\[0.5mm]

\CFARowLabel{Single}
&
\MethodPanel{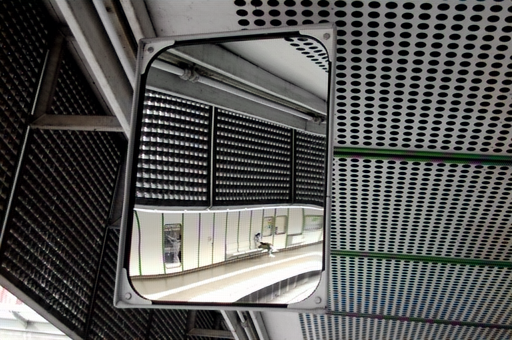}
&
\MethodPanel{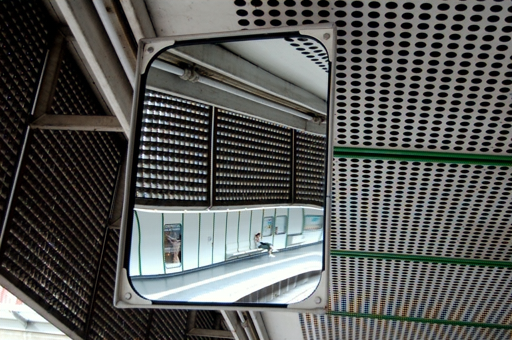}
&
\MethodPanel{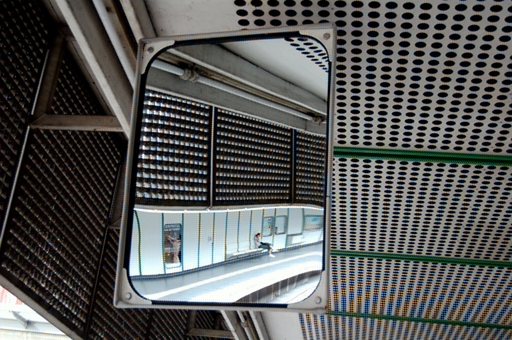}
&
\MethodPanel{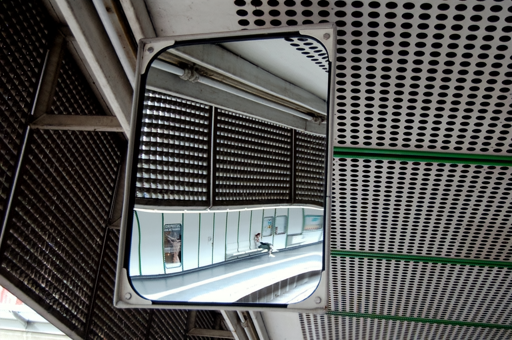}
&
\MethodPanel{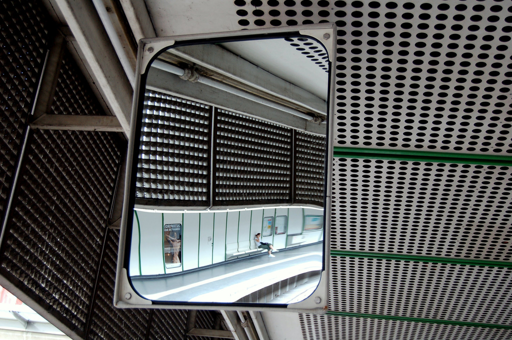}
\\[-0.7mm]

\CFARowLabel{Quad}
&
\MethodPanel{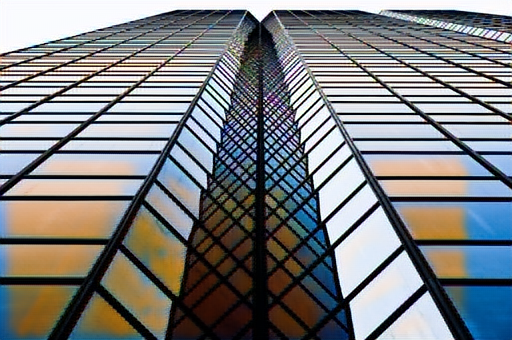}
&
\MethodPanel{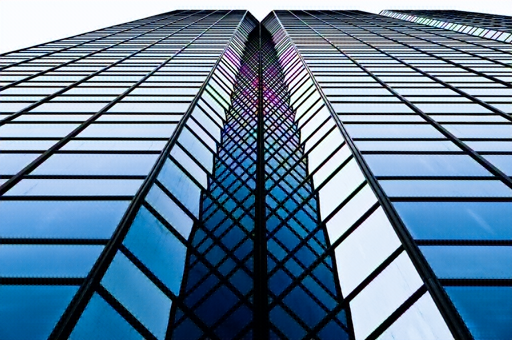}
&
\MethodPanel{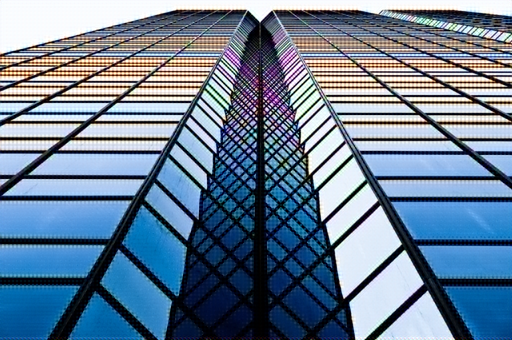}
&
\MethodPanel{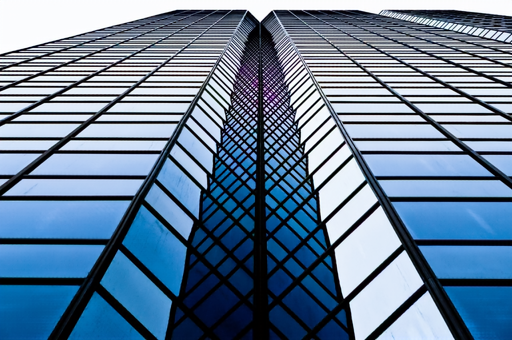}
&
\MethodPanel{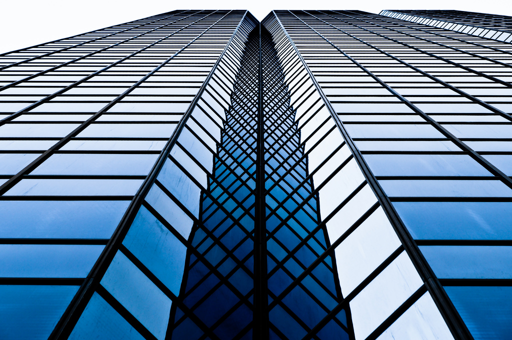}
\\[-0.7mm]

\CFARowLabel{Nona}
&
\MethodPanel{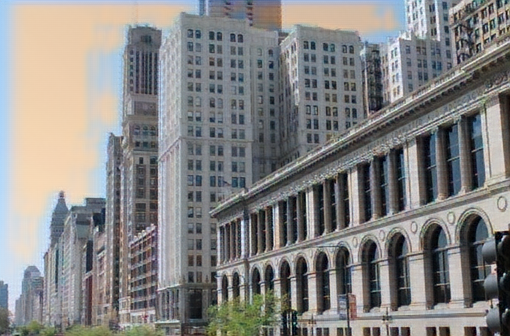}
&
\MethodPanel{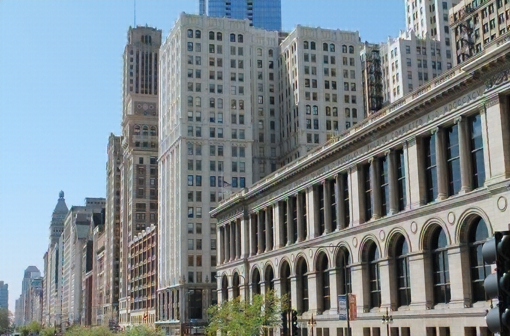}
&
\MethodPanel{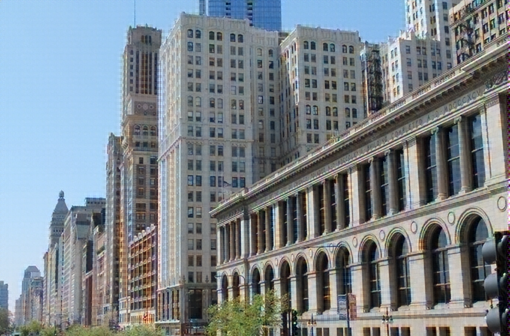}
&
\MethodPanel{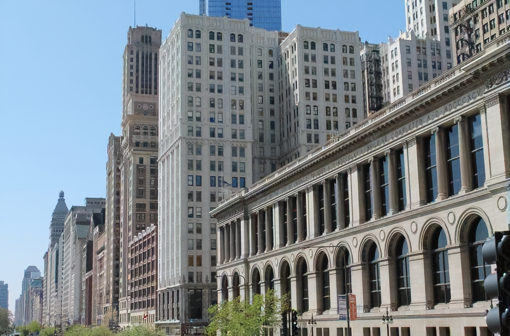}
&
\MethodPanel{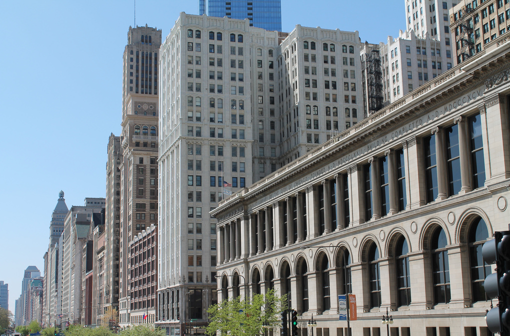}
\\[-0.3mm]

&
\multicolumn{5}{c}{
    \scriptsize\textit{Urban100}~\cite{huang2015single}
}
\\

\end{tabular}

\caption{Additional qualitative comparisons on BSD100, Kodak24, and Urban100 across Single-, Quad-, and Nona-Bayer observations. Results within each row use the same observation, crop, and display range.}
\label{fig:supp_cfa_patterns}
\end{figure*}

Figure~\ref{fig:supp_cfa_patterns} complements the quantitative CFA analysis with fixed-crop comparisons across BSD100, Kodak24, and Urban100 and all three sampling families. These datasets provide natural scenes, photographic detail, and repeated urban structure, respectively. The selected crops emphasize thin boundaries, text-like strokes, and repetitive patterns that are diagnostically difficult for demosaicing; they should therefore be interpreted as hard-case illustrations rather than random-sample estimates.

\begin{figure*}[!t]
\centering
\setlength{\tabcolsep}{0.4pt}
\renewcommand{\arraystretch}{1.0}

\newcommand{\NoisePanel}[1]{%
    \includegraphics[width=\linewidth]{#1}%
}

\newcommand{\NoiseRowLabel}[1]{%
    \rotatebox[origin=c]{90}{%
        \strut\scriptsize\bfseries #1%
    }%
}

\begin{tabular}{
    @{}
    >{\centering\arraybackslash}m{0.028\textwidth}
    @{\hspace{0.4mm}}
    *{4}{>{\centering\arraybackslash}m{0.238\textwidth}}
    @{}
}

&
\scriptsize $\sigma=0$
&
\scriptsize $\sigma=5/255$
&
\scriptsize $\sigma=15/255$
&
\scriptsize $\sigma=25/255$
\\[-0.2mm]

\NoiseRowLabel{Noisy RGB}
&
\NoisePanel{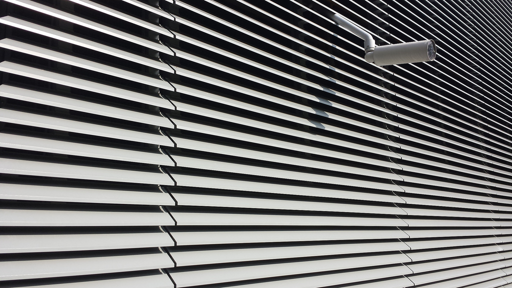}
&
\NoisePanel{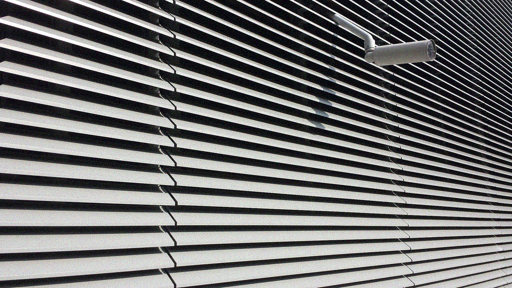}
&
\NoisePanel{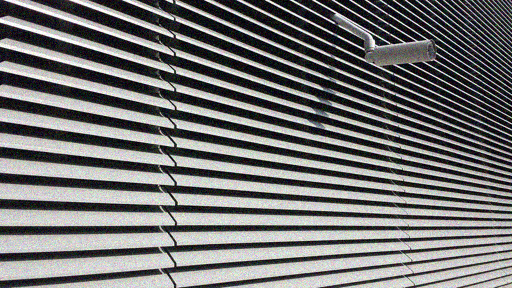}
&
\NoisePanel{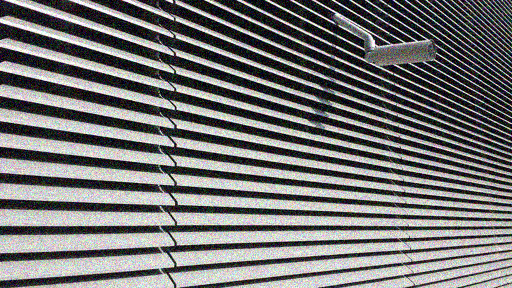}
\\[-0.7mm]

\NoiseRowLabel{Output}
&
\NoisePanel{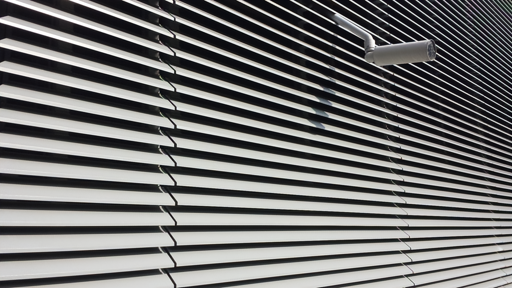}
&
\NoisePanel{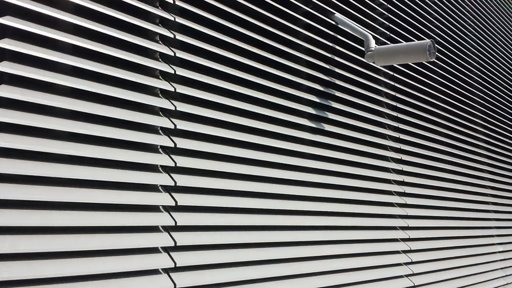}
&
\NoisePanel{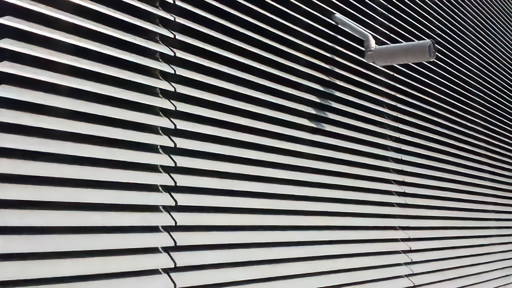}
&
\NoisePanel{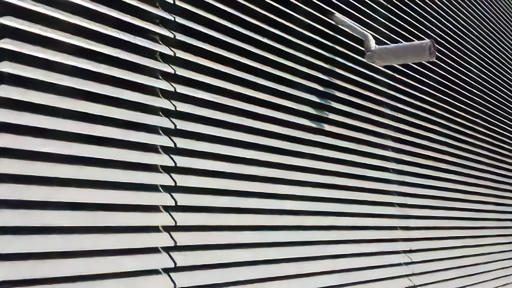}
\\[-0.3mm]

&
\multicolumn{4}{c}{
    \scriptsize\textit{Single-Bayer}
}
\\[0.5mm]

\NoiseRowLabel{Noisy RGB}
&
\NoisePanel{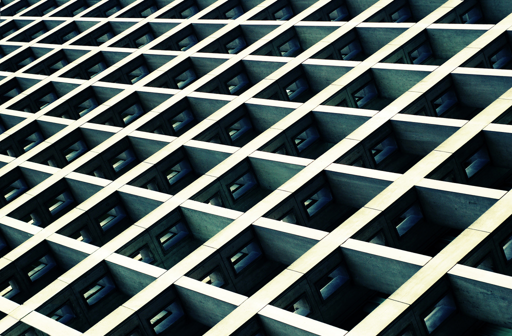}
&
\NoisePanel{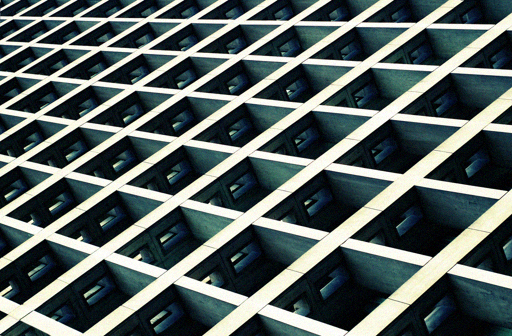}
&
\NoisePanel{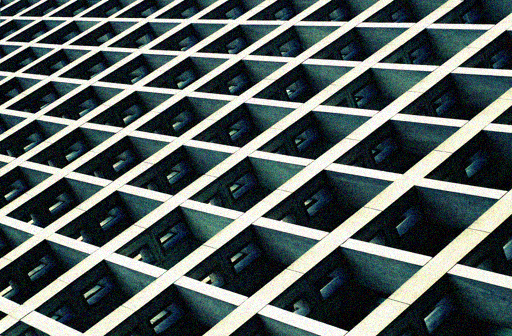}
&
\NoisePanel{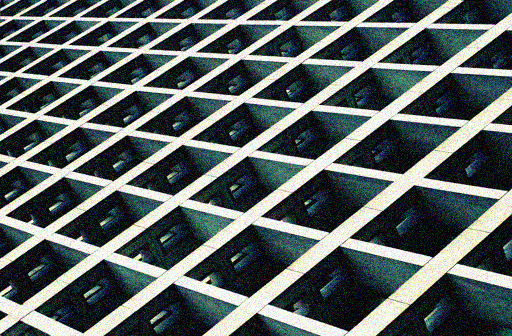}
\\[-0.7mm]

\NoiseRowLabel{Output}
&
\NoisePanel{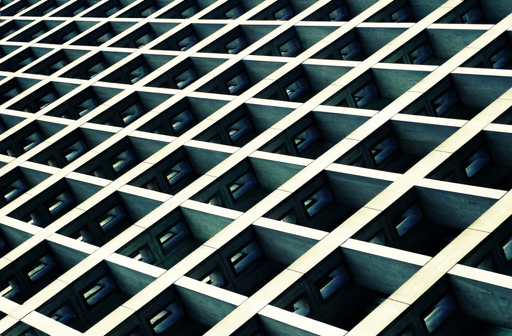}
&
\NoisePanel{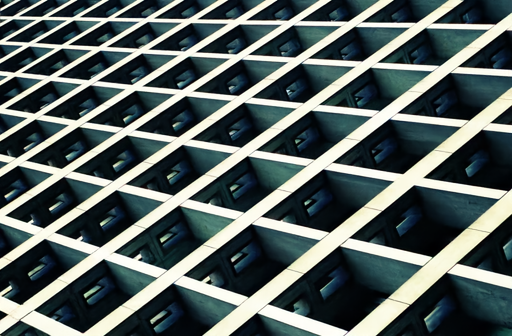}
&
\NoisePanel{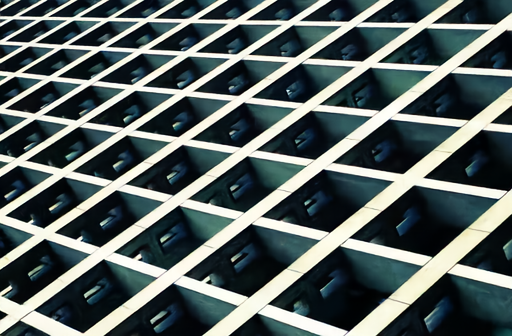}
&
\NoisePanel{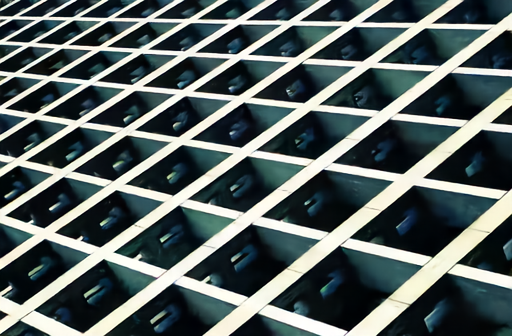}
\\[-0.3mm]

&
\multicolumn{4}{c}{
    \scriptsize\textit{Quad-Bayer}
}
\\[0.5mm]

\NoiseRowLabel{Noisy RGB}
&
\NoisePanel{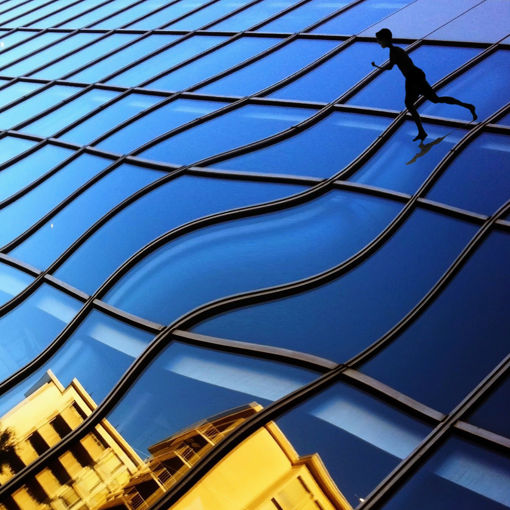}
&
\NoisePanel{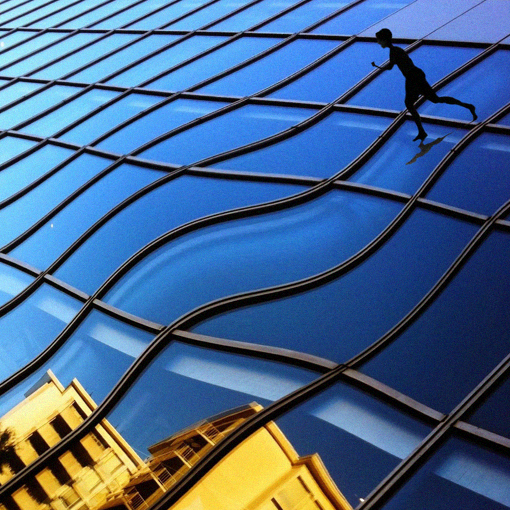}
&
\NoisePanel{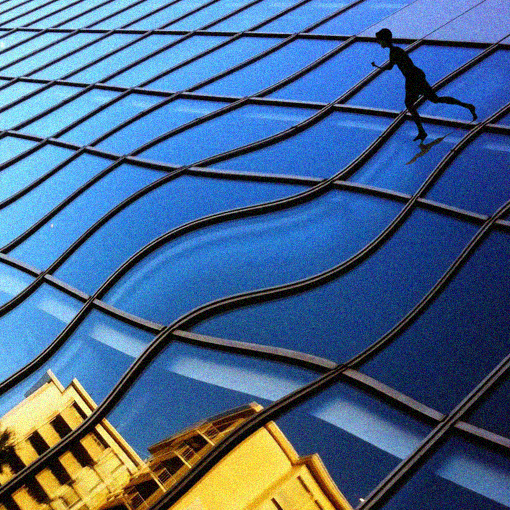}
&
\NoisePanel{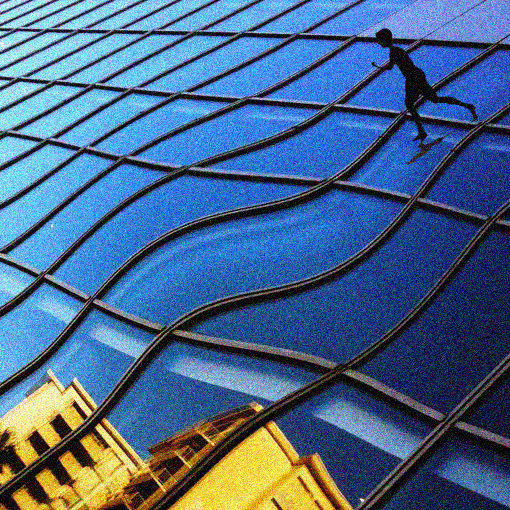}
\\[-0.7mm]

\NoiseRowLabel{Output}
&
\NoisePanel{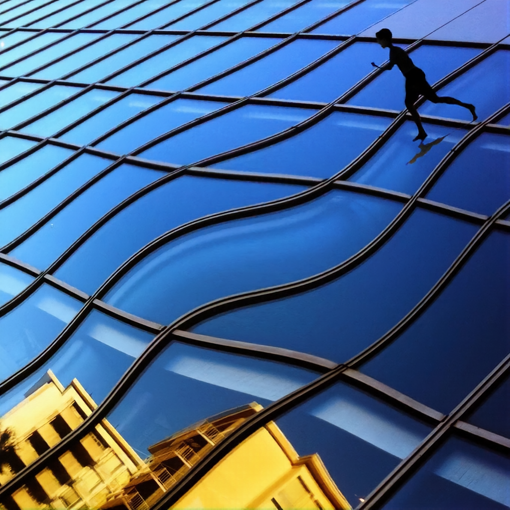}
&
\NoisePanel{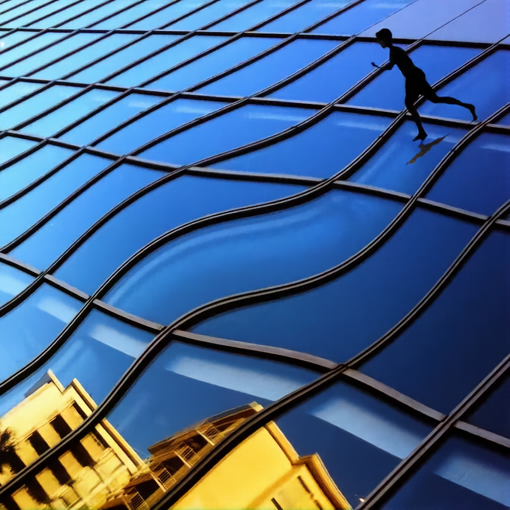}
&
\NoisePanel{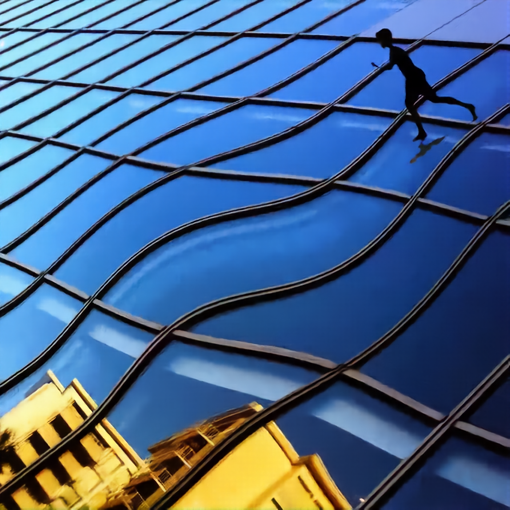}
&
\NoisePanel{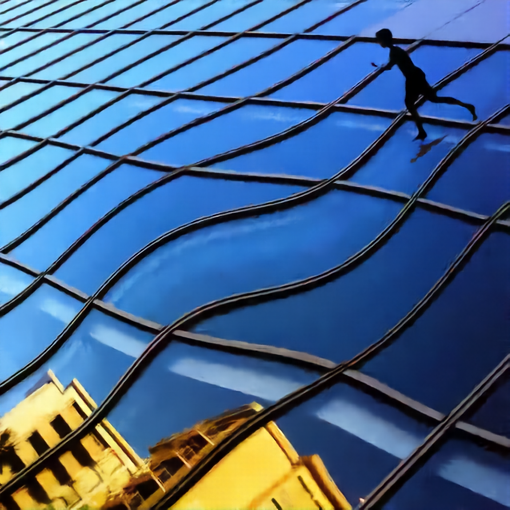}
\\[-0.3mm]

&
\multicolumn{4}{c}{
    \scriptsize\textit{Nona-Bayer}
}
\\

\end{tabular}

\caption{Qualitative restoration results under different noise levels for Single-, Quad-, and Nona-Bayer observations. Each pair of rows shows the noisy RGB images and corresponding outputs at
$\sigma\in\{0,5/255,15/255,25/255\}$.}
\label{fig:supp_noise_comparison}
\end{figure*}

Figure~\ref{fig:supp_noise_comparison} instead holds each scene, CFA family, crop, and model fixed while increasing the noise level, making it possible to inspect degradation from the clean to the strongest-noise condition without changing image content. The primary reading criteria for both figures are edge continuity, preservation of thin strokes and repeated lines, suppression of zippering and false color, residual noise, and over-smoothing of fine texture. These comparisons supplement the dataset-level metrics by showing the spatial form of errors that a single PSNR or SSIM value cannot localize.

\end{document}